\documentclass[aps,prc,preprint,groupedaddress,showpacs]{revtex4-1}
\usepackage{natbib}
\usepackage{graphicx}
\usepackage{amsmath}
\usepackage{hyperref}
\usepackage{color}
\usepackage{bm}

\allowdisplaybreaks

\begin{document}

\preprint{JLAB-THY-17-2515}

\title{Coincidence charged-current neutrino-induced deuteron disintegration for $^2\mathrm{H}_2{^{16}}\mathrm{O}$}

\author{J. W. Van Orden}
\affiliation{Department of Physics, Old Dominion University, Norfolk, VA 23529\\ Jefferson Lab,12000 Jefferson Avenue, Newport News, VA 23606, USA 
\footnote{Notice: Authored by Jefferson Science Associates, LLC under U.S. DOE Contract No. DE-AC05-06OR23177.
The U.S. Government retains a non-exclusive, paid-up, irrevocable, world-wide license to publish or reproduce this manuscript for U.S. Government purposes}
}
\author{T. W. Donnelly}
\affiliation{Center for Theoretical Physics, Laboratory for Nuclear Science and Department of Physics, Massachusetts Institute of Technology, Cambridge, MA 02139, USA}
\author{O. Moreno}
\affiliation{Departamento de Fisica Atomica, Molecular y Nuclear, Facultad de Ciencias Fisicas, Universidad Complutense de Madrid, Ciudad Universitaria, E-28040 Madrid, Spain.}

\begin{abstract}

Semi-inclusive charge-changing neutrino reactions on targets of heavy water are investigated with the goal of determining the relative contributions to the total cross section of deuterium and oxygen in kinematics chosen to emphasize the former. The study is undertaken for conditions where the typical neutrino beam energies are in the few GeV region, and hence relativistic modeling is essential. For this, the previous relativistic approach for the deuteron is employed, together with a spectral function approach for the case of oxygen. Upon optimizing the kinematics of the final-state particles assumed to be detected (typically a muon and a proton) it is shown that the oxygen contribution to the total cross section is suppressed by roughly an order of magnitude compared with the deuterium cross section, thereby confirming that CC$\nu$ studies of heavy water can effectively yield the cross sections for deuterium, with acceptable backgrounds from oxygen. This opens the possibility of using deuterium to determine the incident neutrino flux distribution, to have it serve as a target for which the nuclear structure issues are minimal, and possibly to use deuterium to provide improved knowledge of specific aspects of hadronic structure, such as to explore the momentum transfer dependence of the isovector axial-vector form factor of the nucleon.

\end{abstract}


\maketitle

\section{Introduction}\label{sec:intro}

In two recent studies the subject of {\em semi-inclusive} charge-changing neutrino (CC$\nu$) reactions with nuclei \cite{semi} and application to the special case of deuterium \cite{deut} were presented. Analogous to the semi-inclusive reaction $(e,e'x)$ where one assumes that the scattered electron and some particle $x$ are detected in coincidence, in the weak interaction case one considers reactions of the type $(\nu_\ell, \ell^- x)$ and $(\bar{\nu}_\ell, \ell^+ x)$. These involve incident neutrinos or anti-neutrinos of specific flavor ($\ell = e$, $\mu$ or $\tau$) together with coincident detection of the corresponding charged leptons and some particle $x$. In the present work we shall focus on nucleons ejected from the nucleus, and hence $x=N$, where $N=p$ or $n$. Note that in the nuclear case the ``natural'' type of nucleon may not be the one of interest, whereas for a single-nucleon target and when no other particle is produced other that the final-state nucleon ({\it i.e.,} no pion production, kaon production, {\it etc.}) charge conservation forces the final-state nucleon to be only of one type. Namely, in this latter case one only has reactions of the type $\nu_\ell + n\rightarrow \ell^- + p$ and $\bar{\nu}_\ell + p \rightarrow  \ell^+ + n$. In the present work we shall specialize still further and consider only incident neutrinos, final-state negative leptons and emission of protons ($x=p$). For completeness in defining the terminology commonly being used, we note that reactions where only the final-state leptons are detected, such as $(e,e')$, $(\nu_\ell, \ell^-)$ or $(\bar{\nu}_\ell, \ell^+)$, are called {\em inclusive} reactions.

As has become quite clear in recent years, the typical high-energy neutrino beams used in studies of neutrino oscillations, typically at neutrino energies of around a GeV to tens of GeV, $E_\nu$, have rather broad spreads in energy, which introduces model dependence in the specification of the distance over energy ratio $L/E_\nu$ that enters in the standard oscillation expressions. However, as discussed in \cite{deut}, deuterium provides, at least in principle, an exception to the typical case of heavier nuclei. Namely, once so-called ``no-pion'' events are isolated, all that can occur for the case of incident neutrinos is the reaction $\nu_\ell + {}^2$H$ \rightarrow \ell^- + p + p$. Upon detecting two of the three particles in the final state and knowing the direction of the incident neutrino the neutrino's energy can be reconstructed using nothing beyond the kinematics of the reaction. In \cite{deut} a specific relativistic model for the deuterium ground state and final $NN$ scattering state was employed to model this reaction; in the present study we use the same model for the $A=2$ states and the required electroweak current matrix elements.

This said, there are still practical issues of which to be aware. Namely, making very large target/detectors of hydrogen or deuterium is problematical because of the safety issues involved and the difficulty of providing very large amounts of these nuclei. Using target/detectors of something involving large fractions of deuterium together with other light nuclei, such as heavy water (D$_2$O) or deuterated methane (CD$_4$), might alleviate the safety issue and could provide practical amounts of deuterium, although having other nuclei such as oxygen or carbon present potentially can bring in new considerations. In this study we have focused on a specific case to explore how such mixed nuclear cases behave; specifically, here we consider the case of ${}^2$H$_2 {}^{16}$O. The goal is to take what we have already done for deuterium, add model results for CC$\nu$ semi-inclusive reactions on ${}^{16}$O and determine the degree to which events from the two nuclear species can be separated. One expects the deuteron events to be very peaked and to occur in a different part of the kinematic space involved from the oxygen events, and, as well, the oxygen events to be much more spread in the appropriate kinematic variables so that the ratio of deuterium to oxygen becomes quite favorable. Indeed, we shall show that this is the case.

We will be drawing on our previous study of semi-inclusive CC$\nu$ reactions in \cite{semi} to highlight and quantify the differences of deuterium and a more typical nucleus such as oxygen (here the nucleus could be chosen to be carbon or any other relatively light nucleus). As a specific model for the oxygen case we employ the spectral function approach of \cite{Benhar:1994hw,Benhar:2005dj}. The goal will be to optimize the selection of semi-inclusive events for the case of deuterium and then see what emerges for the ``background'' from the oxygen events.

The paper is organized as follows. In Sect.~\ref{sec:semi} we summarize the necessary formalism for the semi-inclusive CC$\nu$ reaction, taking as a basis the previous study reported in \cite{semi}, and include some of the relevant formalism needed to inter-relate the experimental ``lab frame'' to the so-called ``$q$-frame''. In Sect.~\ref{sec:deut} we specialize the results of the previous section to the case of deuterium to make very clear the advantage provided by this particular nucleus. We do not repeat the discussion of the formalism for the dynamics and currents involved in the deuterium case, since these have been reported in \cite{deut}. For the case of oxygen we present the required formalism in the context of the spectral function in Sect.~\ref{sec:spect}, following which we employ the two models discussed above to obtain typical results for heavy water and present these in Sect.~\ref{sec:results}. In Sect.~\ref{sec:concl} we offer our conclusions, while in the Appendix we collect expressions for the off-shell single-nucleon response functions employed for the oxygen spectral function case.

\section{Semi-Inclusive Cross Section}\label{sec:semi}

\begin{figure}
	\centerline{\includegraphics[height=2in]{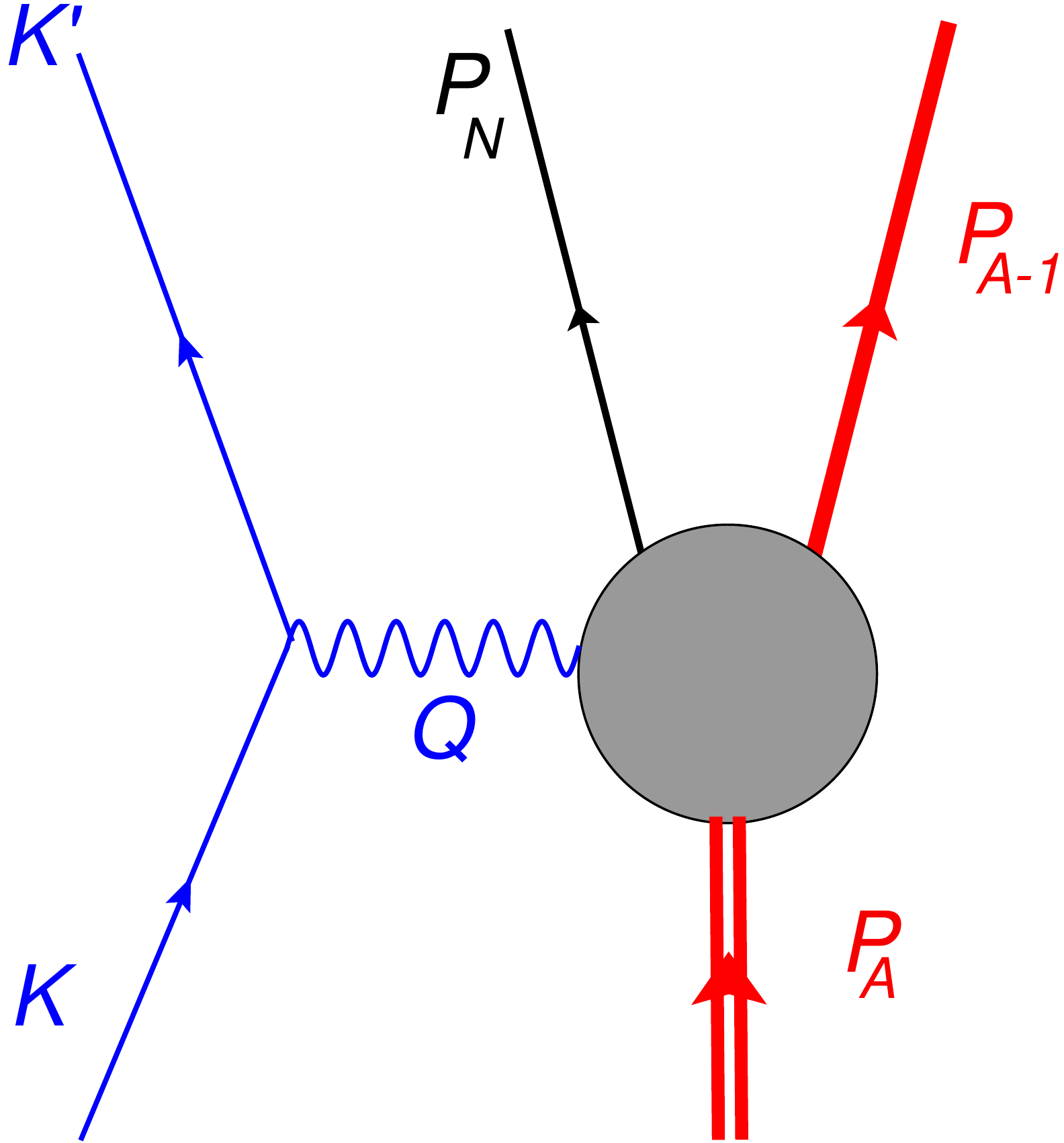}}
	\caption{(color online) Feynman diagram for semi-inclusive charge-changing neutrino reactions involving a target nucleus with nucleon number $A$ with emission and detection of a nucleon with four-momentum $P_N^{\mu}$ together with detection of a final-state charged lepton with four-momentum $K'^{\mu}$ }
	\label{fig:semi}
\end{figure}

Semi-inclusive CC$\nu$ scattering is represented by the Feynman diagram shown in Fig.~\ref{fig:semi}, where $Q^{\mu}$ is the four-momentum of the W-boson,
\begin{equation}
K^\mu=(\varepsilon,\bm{k})
\end{equation}
is the incident lepton four-momentum and
\begin{equation}
{K'}^\mu=(\varepsilon',\bm{k}')
\end{equation}
is the four-momentum of the lepton in the final state,
where $\varepsilon=\sqrt{k^2+m^2}$ and $\varepsilon'=\sqrt{{k'}^2+{m'}^2}$ are the energies of the incident and final leptons with respective masses $m$ and $m'$. Then the four-momentum transfer is
\begin{equation}
Q^\mu=K^\mu-{K'}^\mu=(\varepsilon-\varepsilon',\bm{k}-\bm{k}')=(\omega,\bm{q})\,.
\end{equation}
The four-momentum of the target nucleus with nucleon number $A$ can be written in its rest frame as
\begin{equation}
P_A^\mu=(M_A,\bm{0})\,.
\end{equation}
The four-momentum of the detected nucleon is
\begin{equation}
P_N^\mu=(E_N,\bm{p}_N)\,,
\end{equation}
where  $m_N$ is the nucleon mass, $E_N \equiv \sqrt{\bm{p}_N^2+m_N^2}$ and
the four-momentum of the residual $A-1$ system is
\begin{equation}
P_{A-1}^\mu=(\sqrt{\bm{p}_m^2+W_{A-1}^2},\bm{p}_m)
\end{equation}
with the invariant mass $W_{A-1}$. 

The energy of an incoming neutrino can be determined by measuring the three-momenta of the outgoing charged lepton, which we take to be a muon in what follows (although clearly the $e$ or $\tau$ cases can also be considered), and nucleon, corresponding to kinematics B of \cite{semi}.  In this case the four-fold differential cross section in the laboratory frame is then
\begin{align}
\frac{d\sigma}{dk'd\Omega_{k'}dp_Nd\Omega_N^L}=&\frac{G^2\cos^2\theta_cm_N{k'}^2p_N^2W_{A-1}}{2(2\pi)^5k\varepsilon'E_N}\int \frac{d^3p_m}{\sqrt{p_m^2+W_{A-1}^2}}\eta_{\mu\nu}W^{\mu\nu}\nonumber\\
&\times\delta^4(K+P_A-K'-P_N-P_{A-1})\nonumber\\
=&\frac{G^2\cos^2\theta_cm_N{k'}^2p_N^2W_{A-1}}{2(2\pi)^5k\varepsilon'E_N}\int \frac{d^3p_m}{\sqrt{p_m^2+W_{A-1}^2}}\eta_{\mu\nu}W^{\mu\nu}\nonumber\\
&\times\delta(\varepsilon+M_A-\varepsilon'-E_N-\sqrt{p_m^2+W_{A-1}^2})\delta(\bm{k}-\bm{k}'-\bm{p}_N-\bm{p}_m) \, ,\label{eq:dsig0}
\end{align}
where $G$ is the weak interaction coupling constant and $\theta_c$ is the Cabibbo mixing angle. Defining
\begin{equation}
E_B=\varepsilon'+E_N-M_A
\end{equation}
and
\begin{equation}
\bm{p}_B=\bm{k}'+\bm{p}_N \, ,
\end{equation}
the cross section becomes
\begin{align}
\frac{d\sigma}{dk'd\Omega_{k'}dp_Nd\Omega_N^L}
=&\frac{G^2\cos^2\theta_cm_N{k'}^2p_N^2W_{A-1}}{2(2\pi)^5k\varepsilon'E_N}\int \frac{d^3p_m}{\sqrt{p_m^2+W_{A-1}^2}}\eta_{\mu\nu}W^{\mu\nu}\nonumber\\
&\times\delta(\varepsilon-E_B-\sqrt{p_m^2+W_{A-1}^2})\delta(\bm{k}-\bm{p}_B+\bm{p}_m)\nonumber\\
=&\frac{G^2\cos^2\theta_cm_N{k'}^2p_N^2W_{A-1}}{2(2\pi)^5k\varepsilon'E_N} \frac{1}{\sqrt{(\bm{p}_B-\bm{k})^2+W_{A-1}^2}}\eta_{\mu\nu}W^{\mu\nu}\nonumber\\
&\times\delta(\varepsilon-E_B-\sqrt{(\bm{p}_B-\bm{k})^2+W_{A-1}^2}) \, .
\end{align}
Using the remaining $\delta$-function, the incident neutrino momentum and energy are given by
\begin{equation}
k_0=\frac{1}{a_B}\left(X_Bp_B\cos\theta_B+E_B\sqrt{X_B^2+m^2a_B}\right) \label{Xk0}
\end{equation}
and
\begin{equation}
\varepsilon_0=\frac{1}{a_B}\left(E_BX_B+p_B\cos\theta_B\sqrt{X_B^2+m^2a_B}\right)\,, \label{Xe0}
\end{equation}
where
\begin{equation}
X_B=\frac{1}{2}\left(p_B^2-E_B^2+W_{A-1}^2-m^2\right)
\end{equation}
and
\begin{equation}
a_B=p_B^2\cos^2\theta_B-E_B^2\,.
\end{equation}
The energy-conserving $\delta$-function can be rewritten as
\begin{equation}
\delta(\varepsilon-E_B-\sqrt{(\bm{p}_B-\bm{k})^2+W_{A-1}^2})=\frac{\varepsilon_0\sqrt{(\bm{p}_B-\bm{k})^2+W_{A-1}^2}}{\sqrt{X_B^2+m^2a_B}}\delta(k-k_0)\,.
\end{equation}
The cross section then becomes
\begin{equation}
\frac{d\sigma}{dk'd\Omega_{k'}dp_N^2d\Omega_N^L}
=\frac{G^2\cos^2\theta_cm_N{k'}^2\varepsilon\, p_N^2W_{A-1}v_0}{2(2\pi)^5k\varepsilon'E_N\sqrt{X_B^2+m^2a_B}} \mathcal{F}^2_\chi
\delta(k-k_0) \, ,\label{eq:sigma_lab}
\end{equation}
where $\mathcal{F}^2_\chi \equiv \eta_{\mu\nu}W^{\mu\nu}/v_0$ with $v_0 \equiv (\varepsilon + \varepsilon')^2 - q^2$. The resulting response may be written 
\begin{align}
{\cal F}_\chi^2=&\hat{V}_{CC}(w_{CC}^{VV(I)}+w_{CC}^{AA(I)})
+2\hat{V}_{CL}(w_{CL}^{VV(I)}+w_{CL}^{AA(I)})
+\hat{V}_{LL}(w_{LL}^{VV(I)}+w_{LL}^{AA(I)})\nonumber\\
&+\hat{V}_T(w_{T}^{VV(I)}+w_{T}^{AA(I)})\nonumber\\
&+\hat{V}_{TT}\left[(w_{TT}^{VV(I)}+w_{TT}^{AA(I)})\cos 2\phi_N+(w_{TT}^{VV(II)}+w_{TT}^{AA(II)})\sin 2\phi_N\right]\nonumber\\
&+\hat{V}_{TC}\left[(w_{TC}^{VV(I)}+w_{TC}^{AA(I)})\cos\phi_N
+(w_{TC}^{VV(II)}+w_{TC}^{AA(II)})\sin\phi_N)\right]\nonumber\\
&+\hat{V}_{TL}\left[(w_{TL}^{VV(I)}+w_{TL}^{AA(I)})\cos\phi_N
+(w_{TL}^{VV(II)}+w_{TL}^{AA(II)})\sin\phi_N\right]
\nonumber\\
&+\chi\left[\hat{V}_{T'}w^{VA(I)}_{T'}+\hat{V}_{TC'}(w^{VA(I)}_{TC'}\sin\phi_N
+w_{TC'}^{VA(II)}\cos\phi_N)\right.\nonumber\\
&\left.+\hat{V}_{TL'}(w^{VA(I)}_{TL'}\sin\phi_N
+w^{VA(II)}_{TL'}\cos\phi_N)\right]\label{eq:cal_F2_chi}
\end{align}
with 
\begin{equation}
\chi=\left\{
\begin{array}{rl}
-1 &\mathrm{for\ neutrinos}\\
1  &\mathrm{for\ antineutrinos}
\end{array}\right.\,.
\end{equation}
The kinematic functions $V_a$ and response functions $w^i_j$  are as defined in \cite{semi} with the explicit dependence on azimuthal angle $\phi_N$ defined in the $q$-fixed frame. Response functions labeled by the superscript $(II)$ vanish in the plane-wave limit.

If the neutrino momentum distribution normalized to unity is designated as $P(k)$,
the cross section weighted by this distribution is then given by 
\begin{align}
\left<\frac{d\sigma}{dk'd\Omega_{k'}dp_Nd\Omega_N^L}\right>
=&\int_0^\infty dk \frac{G^2\cos^2\theta_cm_N{k'}^2\varepsilon\, p_N^2W_{A-1}}{2(2\pi)^5k\varepsilon'E_N\sqrt{X_B^2+m^2a_B}} v_0 \mathcal{F}^2_\chi
\delta(k-k_0)P(k)\nonumber\\
=&\frac{G^2\cos^2\theta_cm_N{k'}^2\varepsilon_0\, p_N^2W_{A-1}v_0}{2(2\pi)^5k_0\varepsilon'E_N\sqrt{X_B^2+m^2a_B}} \mathcal{F}^2_\chi
P(k_0)\, .\label{eq:sigma_lab}
\end{align}

\begin{figure}
	\centerline{\includegraphics[height=3in]{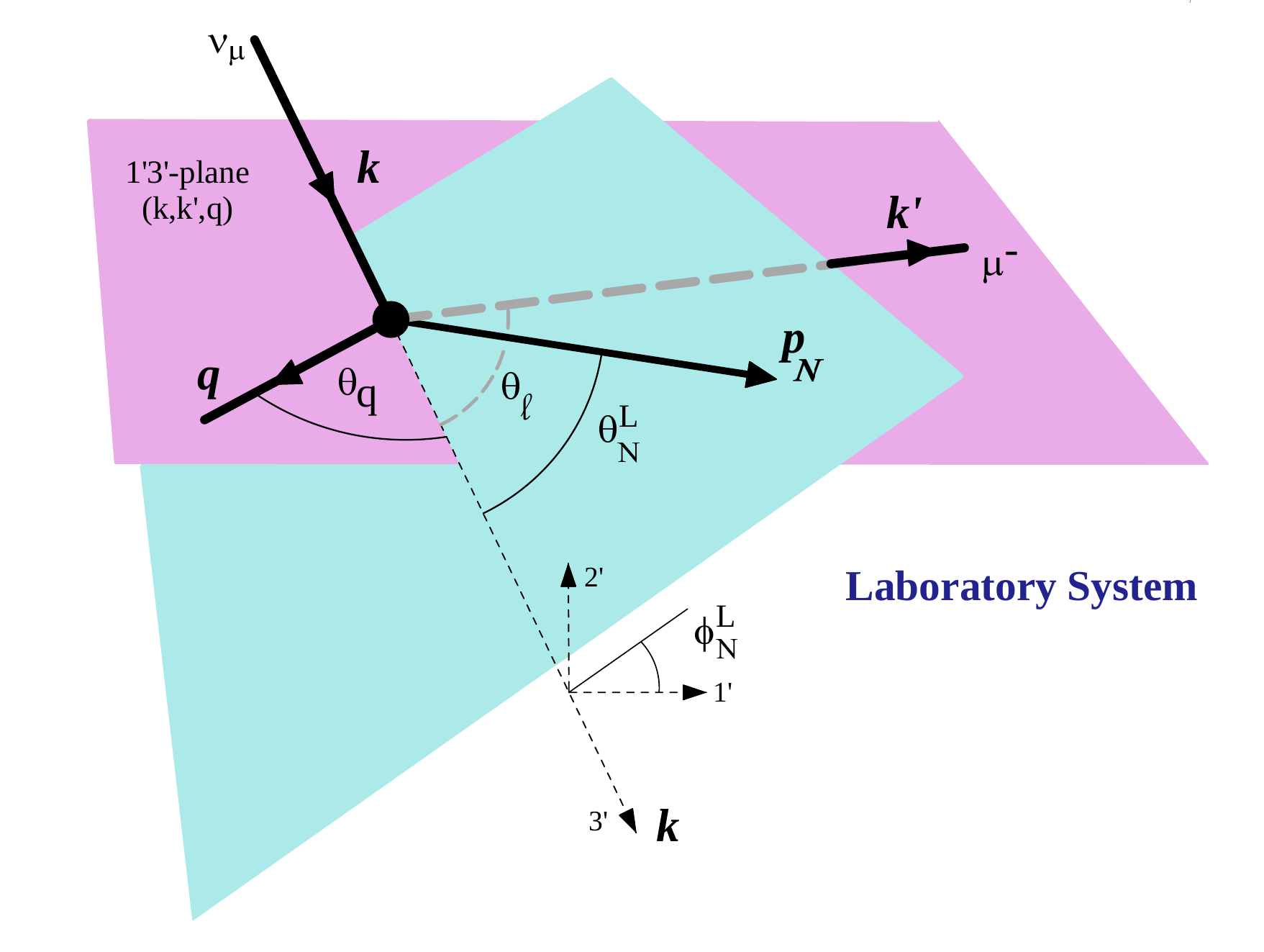}}
	\caption{(color online) Semi-inclusive $(\nu_{\mu},\mu^-p)$ CC$\nu$ reaction in the laboratory frame. Here the incident neutrino with three-momentum ${\bm k}$ is along the $3'$ direction, the neutrino and the final-state muon with three-momentum ${\bm k}'$  lie in the $1'$--$3'$ plane and the normal to the plane defines the $2'$ direction. The outgoing nucleon (here a proton) has three-momentum ${\bm p}_N$ and is traveling in the direction characterized by polar angle $\theta_N^L$ and azimuthal angle $\phi_N^L$ in the lab system, as shown.}
	\label{fig:kinelab}
\end{figure}

\begin{figure}
	\centerline{\includegraphics[height=3in]{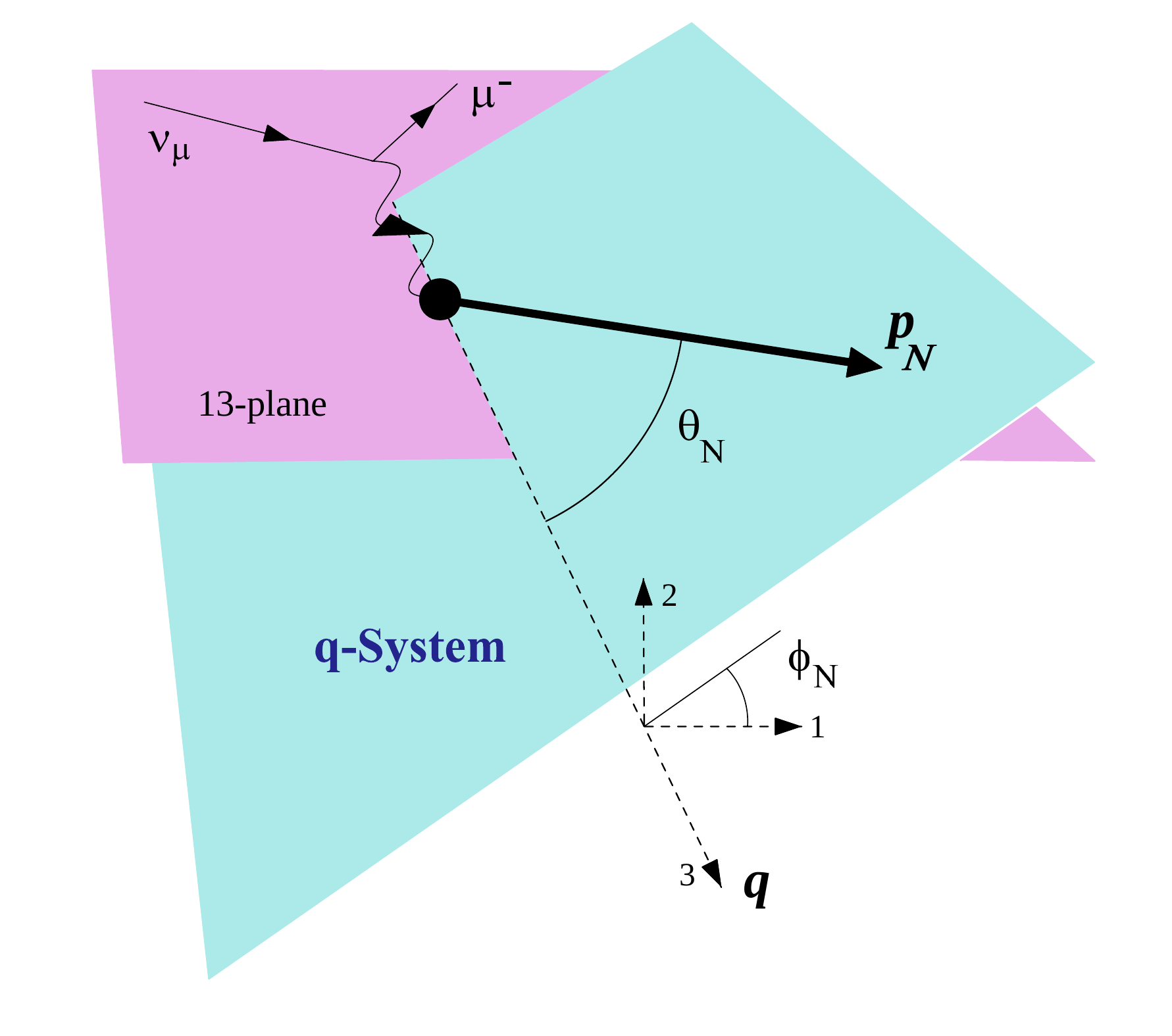}}
	\caption{(color online) Semi-inclusive $(\nu_{\mu},\mu^-p)$ CC$\nu$ reaction in the $q$-system. Here the three-momentum transfer ${\bm q}$ defines the $3$ direction, the neutrino and the final-state muon lie in the $1$--$3$ plane and the normal to the plane defines the $2$ direction. The outgoing nucleon (here a proton) has three-momentum ${\bm p}_N$ and is traveling in the direction characterized by polar angle $\theta_N$ and azimuthal angle $\phi_N$ in the $q$-system, as shown.}
	\label{fig:kineq}
\end{figure}

Next it is useful to inter-relate the variables in the laboratory frame shown in Fig.~\ref{fig:kinelab} to those in the so-called $q$-system shown in Fig.~\ref{fig:kineq}.
We have the following identities relating the angles in the two systems:
\begin{align}
\cos\theta_N=&\cos\theta_N^L \cos\theta_{q}-\cos\phi^L_N \sin\theta_N^L \sin\theta_{q}\label{eq:cos_theta_N}\\
\sin\theta_N=&\sqrt{1-\cos^2\theta_N}\label{eq:sin_theta_N}\\
\cos\phi_N=&\frac{\cos\phi^L_N\sin\theta_N^L\cos\theta_{q}+\cos\theta_N^L\sin\theta_{q}}
{\sin\theta_N}\label{eq:cos_phi_N}\\
\sin\phi_N=&\frac{\sin\phi_N^L\sin\theta_N^L}{\sin\theta_N}\label{eq:sin_phi_N}
 \end{align}
and the inverse relations are given by
\begin{align}
\cos\theta_N^L=&\cos\theta_N \cos\theta_{q}+\cos\phi_N \sin\theta_N \sin\theta_{q}\label{eq:cos_theta_NL}\\
\sin\theta_N^L=&\sqrt{1-\cos^2\theta_N^L}\label{eq:sin_theta_NL}\\
\cos\phi^L_N=&\frac{\cos\phi_N\sin\theta_N\cos\theta_{q}+\cos\theta_N\sin\theta_{q}}{\sin\theta_N^L}\label{eq:cos_phi_NL}\\
\sin\phi^L_N=&\frac{\sin\phi_N\sin\theta_N}{\sin\theta_N^L}\label{eq:sin_phi_NL}\,.
\end{align}
Note that as the neutrino energy changes, even for fixed directions for the outgoing muon and nucleon, the direction of the momentum transfer also changes, and, therefore, through these relationships, the polar and azimuthal angles in the $q$-system also change. The lab system is relevant when experimental issues are being considered; however, the $q$-system with the $3$-direction along the momentum of the exchanged boson has special symmetries that are masked in the lab system. 

Here, we want to express the cross section in lab frame. This can be done by using Eqs.~(\ref{eq:cos_phi_N}) and (\ref{eq:sin_phi_N}) to replace the azimuthal angular dependence in Eq.~(\ref{eq:cal_F2_chi}) and by defining the three-momenta
\begin{equation}
\bm{k}=k\hat{\bm{u}}_{3'}\,,
\end{equation}
\begin{equation}
\bm{k}'=k'\left(\sin\theta_l\hat{\bm{u}}_{1'}+\cos\theta_l\hat{\bm{u}}_{3'}\right)
\end{equation}
and
\begin{equation}
\bm{p}_N=p_N\left(\cos\phi_N^L\sin\theta_N^L\hat{\bm{u}}_{1'}+\sin\phi_N^L\sin\theta_N^L\hat{\bm{u}}_{2'}+\cos\theta_N^L\hat{\bm{u}}_{3'}\right)\,,
\end{equation}
where $\theta_l$ is the lepton scattering angle. The unit vectors in the lab frame are $\hat{\bm{u}}_{1'}$, $\hat{\bm{u}}_{2'}$ and $\hat{\bm{u}}_{3'}$, as shown in Fig.~\ref{fig:kinelab}. The three-momentum transfer is 
\begin{equation}
\bm{q}=\bm{k}-\bm{k}'
\end{equation}
and its square is
\begin{equation}
q^2=k^2+{k'}^2-2kk'\cos\theta_l.
\end{equation}
The angle between $\bm{k}$ and $\bm{q}$ can be obtained from
\begin{equation}
\bm{k}\cdot\bm{q}=kq\cos\theta_{q}=\bm{k}\cdot\bm{k}-\bm{k}\cdot\bm{k}'=k^2-kk'\cos\theta_l \, ,
\end{equation}
which can be solved to yield
\begin{equation}
\cos\theta_{q}=\frac{k-k'\cos\theta_l}{q}\,.
\end{equation}
Similarly we can use
\begin{equation}
\bm{k}\cdot\bm{p}_B=kp_B\cos\theta_B=\bm{k}\cdot(\bm{k}'+\bm{p}_N)=(kk'\cos\theta_l+kp_N\cos\theta_N^L)
\end{equation}
to obtain
\begin{equation}
p_B\cos\theta_B=k'\cos\theta_l+p_N\cos\theta_N^L\,.
\end{equation}
The remaining expressions needed to obtain the cross section in the lab frame are
\begin{equation}
p_B^2={k'}^2+p_N^2+2\bm{k}'\cdot\bm{p}_N={k'}^2+p_N^2+2k'p_N\left(\cos\phi_N^L\sin\theta_N^L\sin\theta_l+\cos\theta_N^L\cos\theta_l\right)
\end{equation}
and
\begin{equation}
p_m^2=k^2+p_B^2-2\bm{k}\cdot\bm{p}_B=k^2+p_B^2-2kp_B\cos\theta_B\,.
\end{equation}

As noted in \cite{deut}, Eq.~(\ref{eq:sigma_lab}) applies also to the case of exclusive scattering from the deuteron by making the substitutions $M_A\rightarrow M_d$ and $W_{A-1}\rightarrow m_N$.

\section{Deuterium}\label{sec:deut}

For the purpose of determining whether the deuterium cross section can be separated from that of oxygen, we wish to choose kinematics which are optimal for the deuteron and then use the values $\bm{k}'$ and $\bm{p}_N$ determined from the deuteron in calculating the semi-inclusive scattering from oxygen.

To obtain the optimal kinematics for scattering from the deuteron we start with Mandelstam $s$ for the virtual $W$ and the deuteron. This is given by
\begin{equation}
s=(P_d+Q)^2=(M_d+\omega)^2-q^2\,.
\end{equation}
The scaling variables \cite{Day:1990mf}
\begin{equation}
y=\frac{(M_d+\omega)\sqrt{s(s-4m_N^2)}}{2s}-\frac{q}{2}
\end{equation}
and
\begin{equation}
Y=y+q
\end{equation}
can be used to obtain limiting values for the magnitude of the missing momentum $p_m$ as
\begin{equation}
|y|\leq p_m\leq Y\,.
\end{equation}

Since the deuteron cross section behaves roughly as the deuteron momentum distribution $n(p_m)$, which peaks at $p_m=0$, the cross section can be optimized by choosing kinematics such that $y=0$. Solving this for the incident neutrino energy yields
\begin{align}
\varepsilon_0=&\frac{1}{2
	\left[(\varepsilon'-M_d+m_N)^2-{k'}^2 \cos^2(\theta_l)\right]}\left\{\zeta k' \cos(\theta_l)\left[ -2
{\varepsilon'}^2 \left(m^2-2
(M_d-m_N)^2\right)\right.\right.\nonumber\\
&\left.\left.-4\varepsilon' (M_d-m_N)\left(-m^2+M_d^2-2 M_d m_N+{m'}^2\right)+2
{k'}^2 m^2 \cos (2\theta_l)+m^4-2 m^2 M_d^2\right.\right.\nonumber\\	
&\left.\left.+4 m^2 M_d
m_N-4 m^2
m_N^2+M_d^4-4 M_d^3
m_N+4 M_d^2 m_N^2+2
M_d^2 {m'}^2-4 M_d
m_N
{m'}^2+{m'}^4\right]^{\frac{1}{2}}\right.\nonumber\\
&\left.-2
{\varepsilon'}^2 M_d+2
{\varepsilon'}^2 m_N+\varepsilon'
m^2+3 \varepsilon' M_d^2-6
\varepsilon' M_d m_N+2
\varepsilon' m_N^2+\varepsilon'
{m'}^2-m^2 M_d+m^2
m_N\right.\nonumber\\
&\left.-M_d^3+3 M_d^2
m_N-2 M_d
m_N^2-M_d
{m'}^2+m_N {m'}^2\right\} \, ,
\end{align}
where
\begin{equation}
\zeta=\left\{
\begin{array}{rr}
-1 & \mathrm{for}\ \theta_l\leq \frac{\pi}{2}\\
1 &  \mathrm{for}\ \theta_l> \frac{\pi}{2}
\end{array}\right.
\end{equation}

Four-momentum conservation for the deuteron requires that
\begin{align}
0=&M_d+\omega-\sqrt{p_N^2+m_N^2}-\sqrt{p_m^2+m_N^2}\label{eqn:deuteron_energy}\\
\bm{0}=&\bm{q}-\bm{p}_N+\bm{p}_m \, .\label{eqn:deuteron_momentum}
\end{align}
Using Eq.~(\ref{eqn:deuteron_energy}) the square of the detected nucleon momentum is
\begin{equation}
p_N^2=\left(M_d+\omega-\sqrt{p_m^2+m_N^2}\right)^2-m_N^2\,.\label{eq:p_N_squared}
\end{equation}
Using Eq.~(\ref{eqn:deuteron_momentum}),
\begin{equation}
\bm{p}_m=\bm{q}-\bm{p}_N\,,
\end{equation}
yields
\begin{equation}
p_{m}^2=q^2+p_N^2-2p_Nq\cos\theta_N\,.
\end{equation}
Solving this for $\cos\theta_N$ gives
\begin{equation}
\cos\theta_N=\frac{q^2+p_N^2-p_m^2}{2p_Nq}\,.\label{eq:cos_theta_N_2}
\end{equation}
By specifying $k'$, $\theta_l$, $\phi_N$ and using Eqs.~(\ref{eq:p_N_squared}) and (\ref{eq:cos_theta_N_2}), the lab frame angles are then given by Eqs.~(\ref{eq:cos_theta_NL}), (\ref{eq:sin_theta_NL}), (\ref{eq:cos_phi_NL}) and (\ref{eq:sin_phi_NL}). This provides a complete set of input variables to evaluate the deuteron and oxygen cross sections. Note that Eq.~(\ref{eq:cos_theta_N_2}) results in a correlation of the values of $p_N$ and $\cos\theta^L_N$.

All of the conditions required by these constrained kinematics can only be satisfied by limiting
\begin{equation}
0\leq\theta_l\leq\left\{
\begin{array}{ll}
\cos^{-1}\left(\frac{\varepsilon'-M_d+m_N}{k'}\right)& \mathrm{for}\ -k'<\varepsilon'-M_d+m_N\leq k'\\
\pi & \mathrm{for}\ \varepsilon'-M_d+m_N\leq -k'
\end{array}
\right.\,.\label{eq:theta_l_max}
\end{equation}

The deuterium matrix elements needed to construct the cross section are described in \cite{deut}.

\section{Spectral Function}\label{sec:spect}

\begin{figure}
	\centerline{\includegraphics[height=2in]{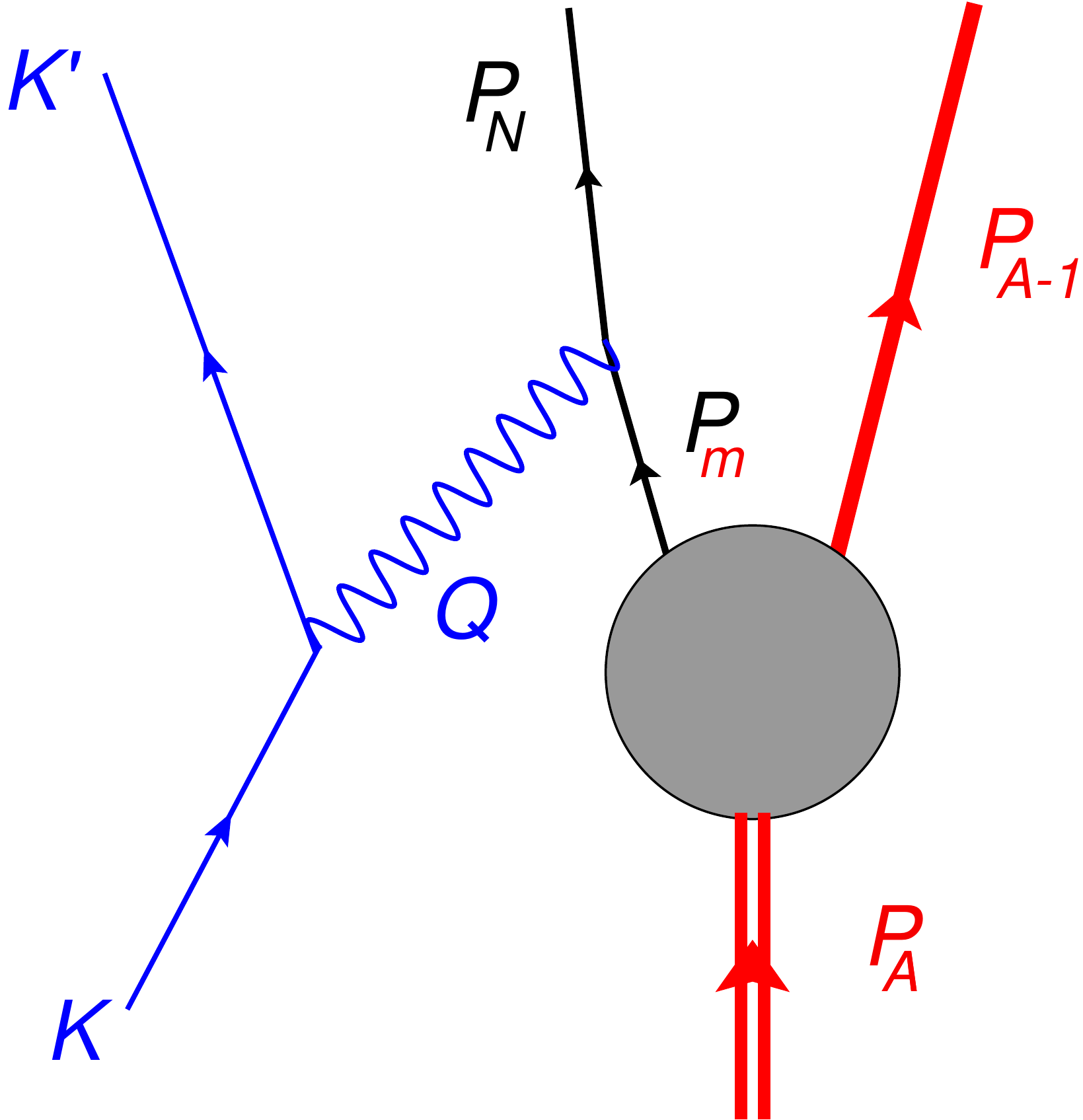}}
	\caption{(color online) Feynman diagram for a factorized approximation to the semi-inclusive charge-changing neutrino reaction illustrated for the general case in Fig.~1. }
	\label{fig:PWIA_semi}
\end{figure}

For this work we estimate the oxygen semi-inclusive cross sections using a factorized spectral function model. The current matrix element for this model can be written as
\begin{equation}
\left<\bm{p}_N,s_N;\bm{P}_{A-1},s_{A-1}\left|J^\mu(q)\right|\bm{P}_A,s_A\right>=\bar{u}(\bm{p}_N,s_N)_a J^\mu(q)_{ab}\Psi(P_{A-1},s_{A-1};P_A,s_A)_{bc} \, ,
\end{equation}
where $s_N$, $s_A$ and $s_{A-1}$ are the spins of the ejected proton, target nucleus and residual system, respectively, and $\Psi(P_{A-1},s_{A-1};P_A,s_A)$ represents a three-point function with the $A$ line truncated. The Dirac indices are explicitly indicated. The nuclear response tensor is then given by
\begin{align}
W^{\mu\nu}=&\sum_{s_N}\sum_{s_A}\sum_{s_{A-1}}
\bar{u}(\bm{p}_N,s_N)_a J^\nu(q)_{ab}\Psi(P_{A-1},s_{A-1};P_A,s_A)_{bc}\nonumber\\
&\times\bar{\Psi}(P_{A-1},s_{A-1};P_A,s_A)_{cd} J^\mu(-q)_{de}u(\bm{p}_N,s_N)_e\nonumber\\
=&\sum_{s_N}
\bar{u}(\bm{p}_N,s_N)_a J^\nu(q)_{ab}\frac{1}{8\pi}\Lambda^+(\bm{p}_m)_{bd}S(p_m,E_m) J^\mu(-q)_{de}u(\bm{p}_N,s_N)_e\nonumber\\
=&\frac{1}{8\pi} \mathrm{Tr}\left[J^\mu(-q)\Lambda^+(\bm{p}_N)J^\nu(q)\Lambda^+(\bm{p}_m)\right]S(p_m,E_m)\nonumber\\
=&\frac{1}{8\pi} w^{\mu\nu}(P_{A}-P_{A-1},Q)S(p_m,E_m)\,, 
\end{align}
where $w^{\mu\nu}(P_{A}-P_{A-1},Q)$ is an off-shell single-nucleon response tensor and $S(p_m,E_m)$ is the spectral function.
The missing energy is approximated by
\begin{equation}
E_m\cong E_s+\mathcal{E} \, ,
\end{equation}
where $E_s$ is the separation energy,
\begin{equation}
\mathcal{E}=\sqrt{p_m^2+W_{A-1}^2}-\sqrt{p_m^2+{W_{A-1}^0}^2}\,,
\end{equation} 
and $W_{A-1}^0$ is the invariant mass of the lowest state of the residual system. Energy conservation requires that
\begin{align}
0=&M_A+\omega-\sqrt{p_N^2+m_N^2}-\sqrt{p_m^2+W_{A-1}^2}\nonumber\\
=&M_A+\omega-\sqrt{p_N^2+m_N^2}-\sqrt{p_m^2+W_{A-1}^2}+\sqrt{p_m^2+{W_{A-1}^0}^2}-\sqrt{p_m^2+{W_{A-1}^0}^2}\nonumber\\
=&M_A+\omega-\sqrt{p_N^2+m_N^2}-\mathcal{E}-\sqrt{p_m^2+{W_{A-1}^0}^2}\,.
\end{align}
So $\mathcal{E}$ can also be written as
\begin{equation}
\mathcal{E}=M_A+\omega-\sqrt{p_N^2+m_N^2}-\sqrt{p_m^2+{W_{A-1}^0}^2}\,.
\end{equation}
From momentum conservation $\bm{p}_N=\bm{q}-\bm{p}_m$, and therefore
\begin{equation}
\mathcal{E}=M_A+\omega-\sqrt{(\bm{q}-\bm{p}_m)^2+m_N^2}-\sqrt{p_m^2+{W_{A-1}^0}^2}\,.
\end{equation}
The range of $\mathcal{E}$ is then limited by
\begin{equation}
\mathcal{E}_+\leq\mathcal{E}\leq\mathcal{E}_-\,,
\end{equation}
where
\begin{equation}
\mathcal{E}_-=M_A+\omega-\sqrt{(p_m-q)^2+m_N^2}-\sqrt{p_m^2+{W_{A-1}^0}^2}
\end{equation} 
and
\begin{equation}
\mathcal{E}_+=\max(M_A+\omega-\sqrt{(p_m+q)^2+m_N^2}-\sqrt{p_m^2+{W_{A-1}^0}^2},0)\,.
\end{equation}

The normalization of the spectral function $S(p_m,E_m)$ is defined here such that
\begin{equation}
\int_0^\infty dE_m S(p_m,E_m)=n(p_m)
\end{equation}
is the momentum distribution and
\begin{equation}
\frac{1}{(2\pi)^3}\int_0^\infty dp_m p_m^2 n(p_m)=A-Z\,.
\end{equation}

Expressing the four-momentum of the struck nucleon as
\begin{equation}
P_A^\mu-P_{A-1}^\mu=(M_A-\sqrt{p_m^2+W_{A-1}^2},-\bm{p}_m)\,,
\end{equation}
defining
\begin{equation}
\bm{p}=-\bm{p}_m
\end{equation}
and using energy conservation
\begin{equation}
M_A-\sqrt{p^2+W_{A-1}^2}=\sqrt{p_N^2+m_N^2}-\omega\,,
\end{equation}
one has
\begin{align}
P_A^\mu-P_{A-1}^\mu=&(\sqrt{p_N^2+m_N^2}-\omega,\bm{p})\nonumber\\
=&(\sqrt{p_N^2+m_N^2}-\omega-\sqrt{p^2+m_N^2}+\sqrt{p^2+m_N^2},\bm{p})\nonumber\\
=&(\sqrt{p_N^2+m_N^2}-\omega-\sqrt{p^2+m_N^2},\bm{0})+(\sqrt{p^2+m_N^2},\bm{p})\nonumber\\
=&(\delta,\bm{0})+(\sqrt{p^2+m_N^2},\bm{p})=\Delta^\mu+P^\mu \, ,
\end{align}
where
\begin{equation}
P^{\mu}=(\sqrt{p^2+m_N^2},\bm{p})
\end{equation}
is an on-shell four-vector and 
\begin{equation}
\Delta^\mu=(\delta,\bm{0})
\end{equation}
is off-shell with
\begin{equation}
\delta=\sqrt{p_N^2+m_N^2}-\sqrt{p^2+m_N^2}-\omega\,.
\end{equation}

The quantity $\mathcal{F}^2_\chi$ in Eq.~(\ref{eq:sigma_lab}) is then given by
\begin{equation}
\mathcal{F}^2_\chi\cong\frac{1}{8\pi}\widetilde{\mathcal{F}}^2_\chi S(p_m,E_m) \, ,
\end{equation}
where
\begin{align}
\widetilde{\mathcal{F}}^2_\chi=&  \widehat{V}_{CC}\left(\widetilde{w}^{VV(I)}_{CC}+\widetilde{w}^{AA(I)}_{CC}\right)+2\widehat{V}_{CL}\left(\widetilde{w}^{VV(I)}_{CL}+\widetilde{w}^{AA(I)}_{CL}\right)+\widehat{V}_{LL}\left(\widetilde{w}^{VV(I)}_{LL}+\widetilde{w}^{AA(I)}_{LL}\right)\nonumber\\
&+\widehat{V}_{T}\left(\widetilde{w}^{VV(I)}_{T}+\widetilde{w}^{AA(I)}_{T}\right)
 +\widehat{V}_{TT}\left(\widetilde{w}^{VV(I)}_{TT}+\widetilde{w}^{AA(I)}_{TT}\right)\cos 2\phi_N  \nonumber\\
 &+\widehat{V}_{TC}\left(\widetilde{w}^{VV(I)}_{TC}+\widetilde{w}^{AA(I)}_{TC}\right)\cos\phi_N+\widehat{V}_{TL}\left(\widetilde{w}^{VV(I)}_{TL}+\widetilde{w}^{AA(I)}_{TL}\right)\cos\phi_N \nonumber\\
& 
+\chi \left[ \widehat{V}_{T^{\prime }}\widetilde{w}^{VA(I)}_{T^{\prime }}+\widehat{V}_{TC^{\prime }}\widetilde{w}^{VA(I)}_{TC^{\prime }}\sin\phi_N+\widehat{V}_{TL^{\prime }}\widetilde{w}^{VA(I)}_{TL^{\prime }}\sin\phi_N\right] \, .
\end{align}
The off-shell single-nucleon response functions $\widetilde{w}^i_j$ are listed in the Appendix.

Since the invariant mass of the residual $A-1$ system is not measured, it is necessary that the semi-inclusive cross section be integrated over all possible values of $W_{A-1}$ to give
\begin{align}
\left<\frac{d\sigma}{dk'd\Omega_{k'}^Ldp_Nd\Omega_N^L}\right>
=&\int_{W^0_{A-1}}^{\infty}dW_{A-1}\frac{G^2\cos^2\theta_cm_N{k'}^2\varepsilon_0 p_N^2W_{A-1}v_0}{8(2\pi)^6k_0\varepsilon'E_N\sqrt{X_B^2+m^2a_B}} \widetilde{\mathcal{F}}^2_\chi S(p_m,E_m)
P(k_0) \, ,
\end{align}
where $W^0_{A-1}$ is the lowest possible mass for the residual system which in some cases may not be a bound state. For the specific case considered in the present study this corresponds to the ground-state mass of $^{15}$O.

Note that the integral over the invariant mass requires that $k_0$ and $\varepsilon_0$ in Eqs.~(\ref{Xk0}) and (\ref{Xe0}) must take on a range of values rather than being fixed as in the case of the deuteron.

\section{Results}\label{sec:results}

\begin{figure}
	\centerline{\includegraphics[height=3in]{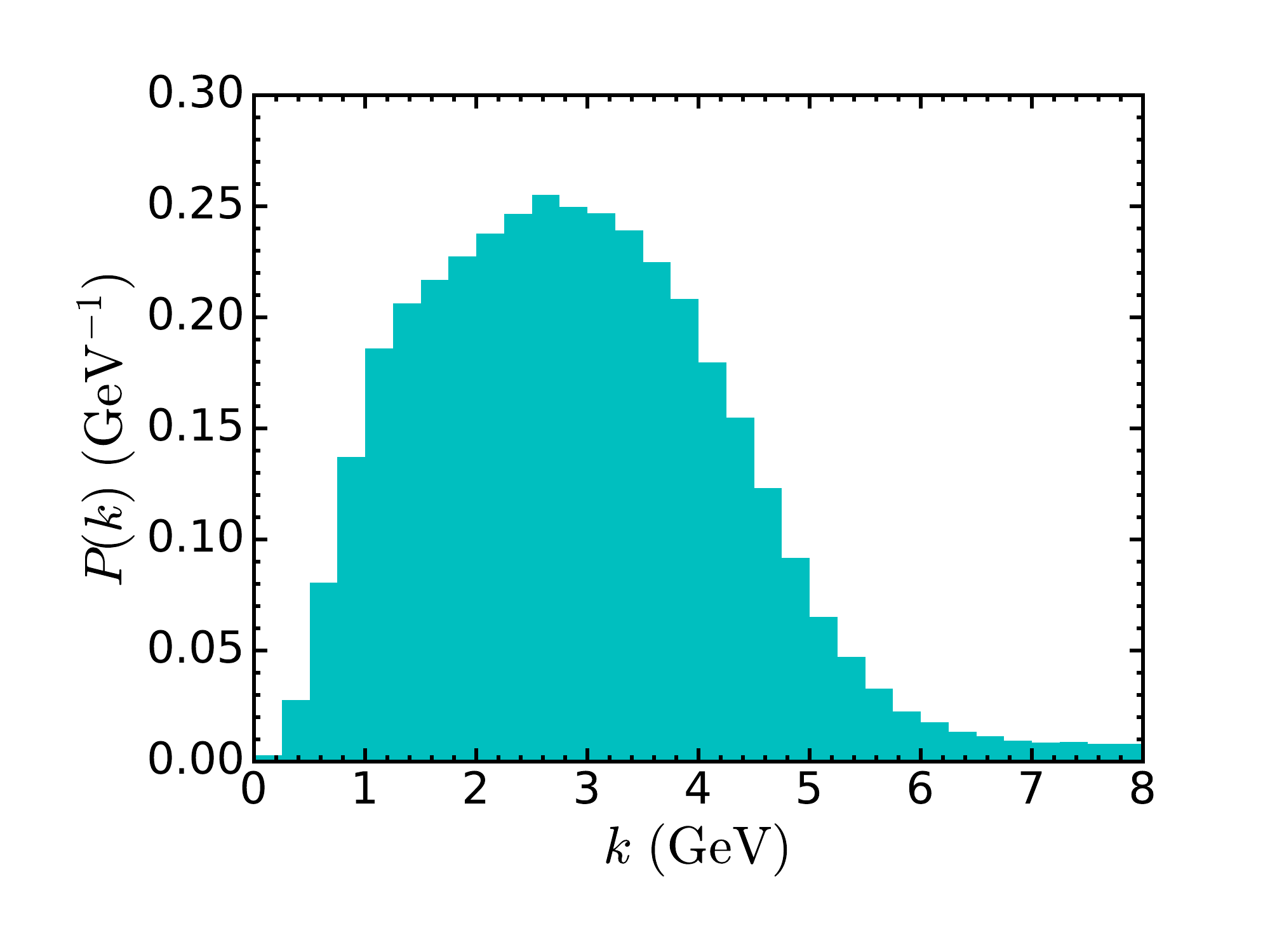}}	\caption{(color online) DUNE flux converted to a probability density as a function of $k$ in GeV.  }\label{fig:DUNE_flux}
\end{figure}

For the purposes of this paper, we have chosen to weight the cross sections using the flux momentum distribution for the DUNE experiment \cite{Alion:2016uaj} normalized to unit area, represented $P(k)$ as shown in Fig. \ref{fig:DUNE_flux}. The spectral function for oxygen is from \cite{Benhar:1994hw,Benhar:2005dj} renormalized according to the units and conventions used here.

\begin{figure}
	\centerline{\includegraphics[height=2in]{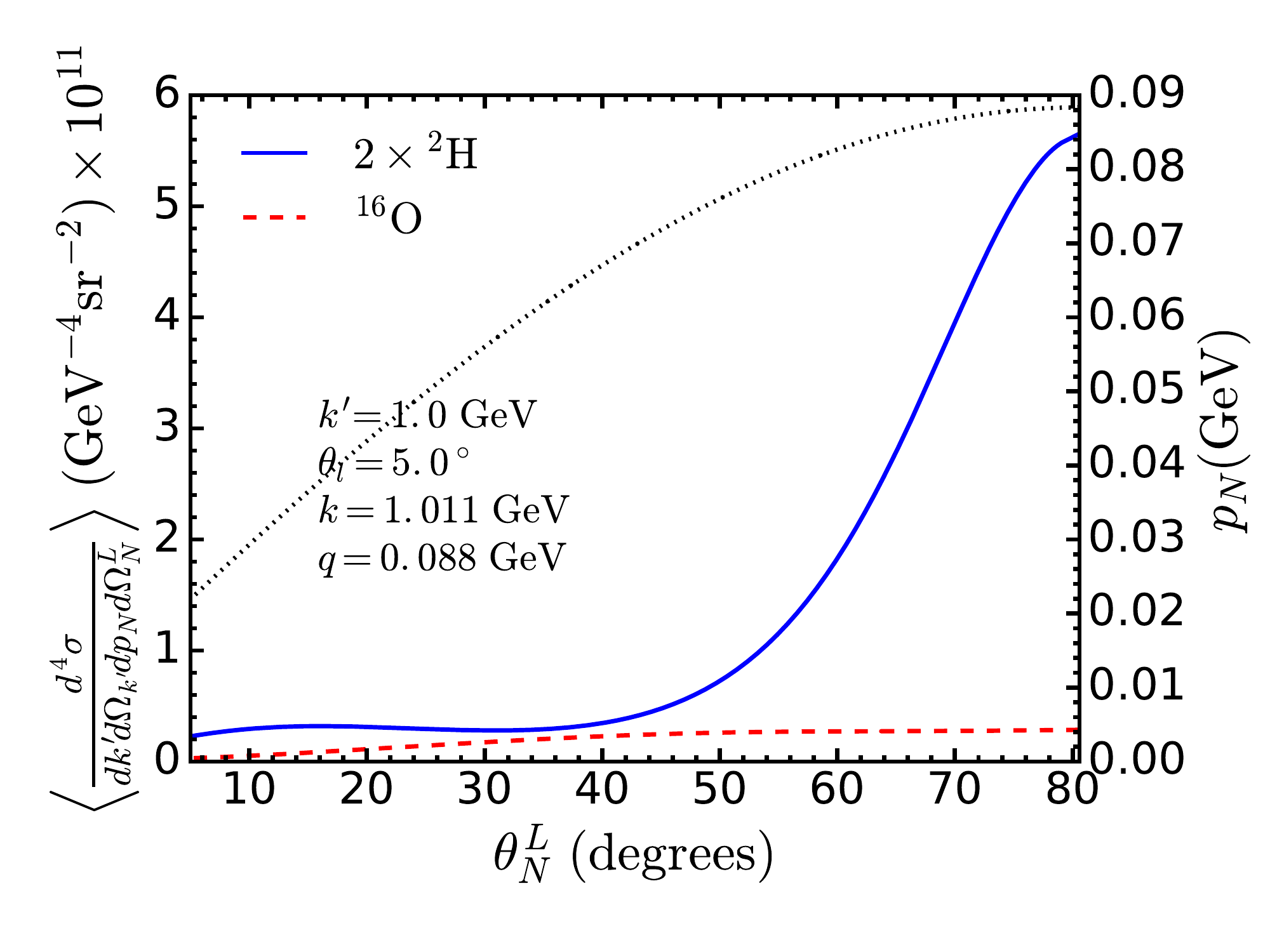}\includegraphics[height=2in]{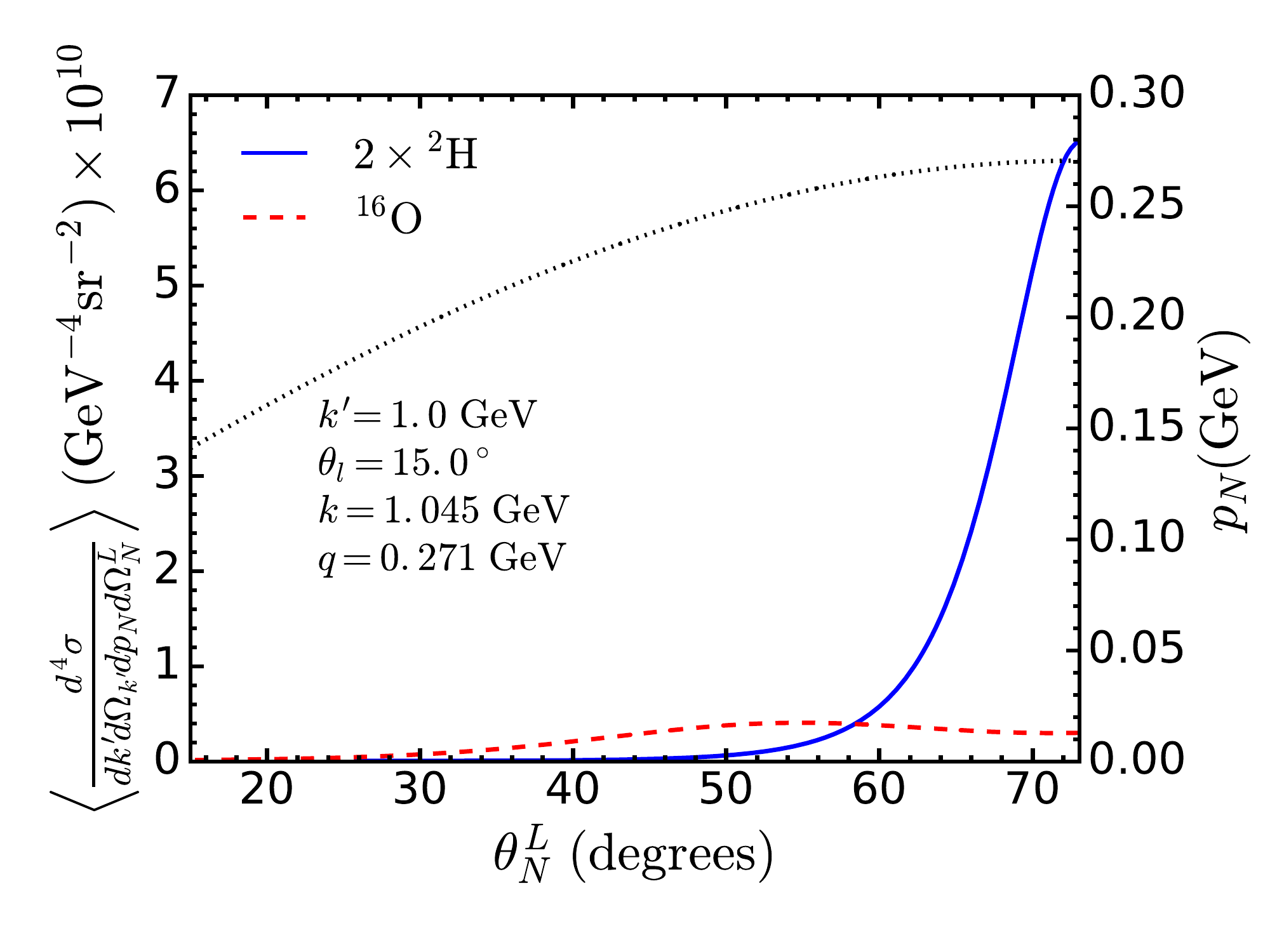}}
	\centerline{\includegraphics[height=2in]{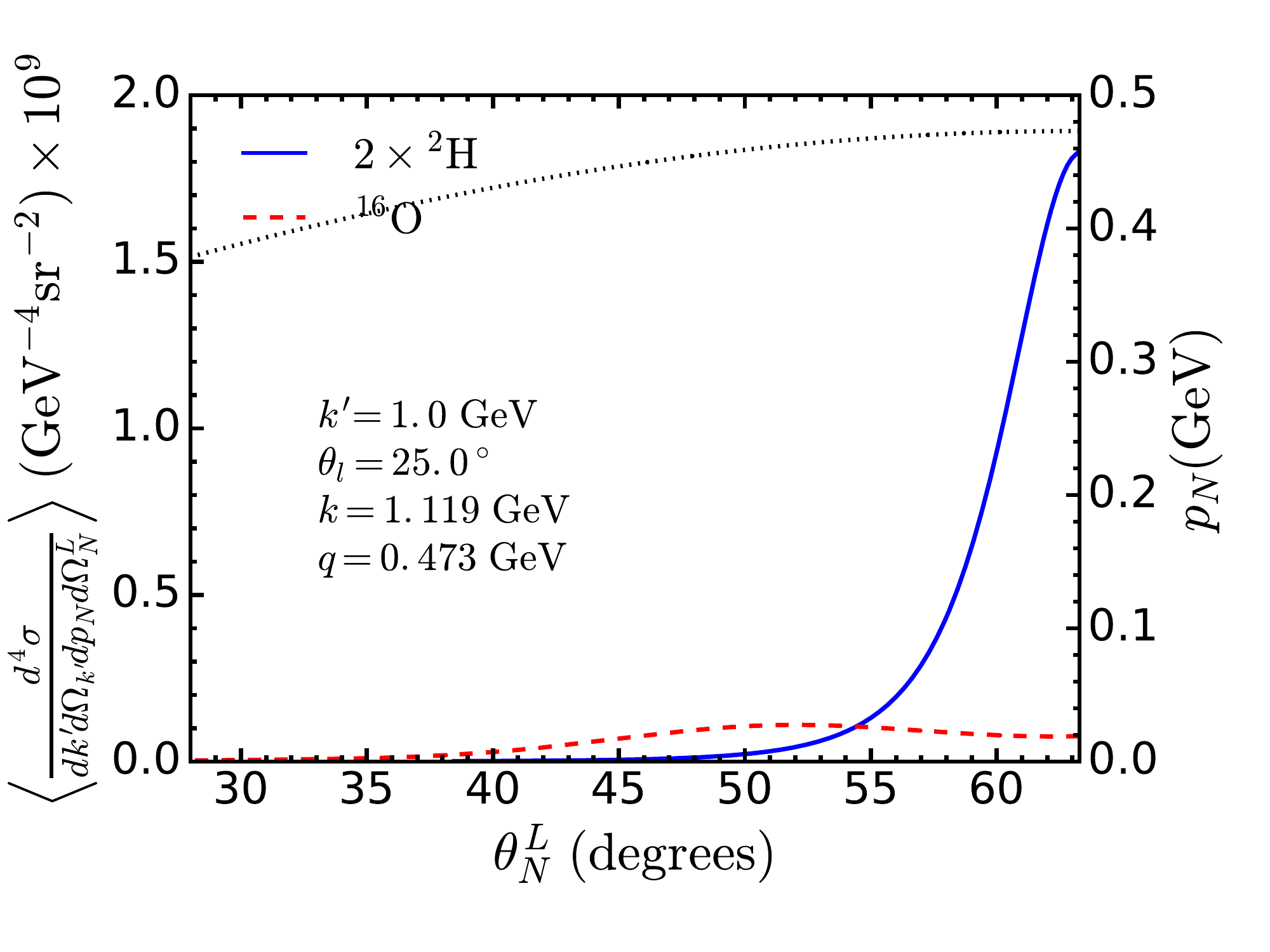}\includegraphics[height=2in]{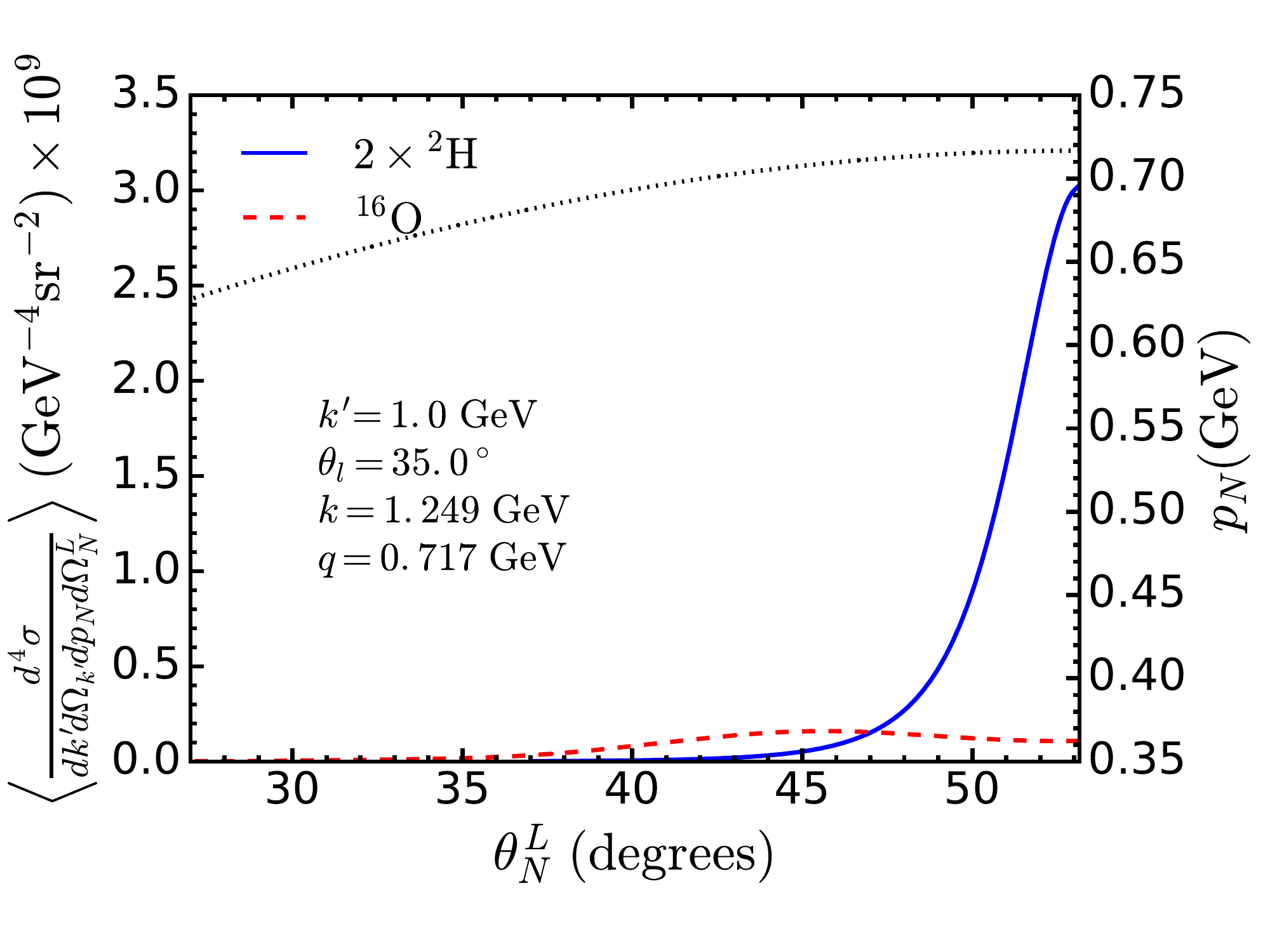}}
	\centerline{\includegraphics[height=2in]{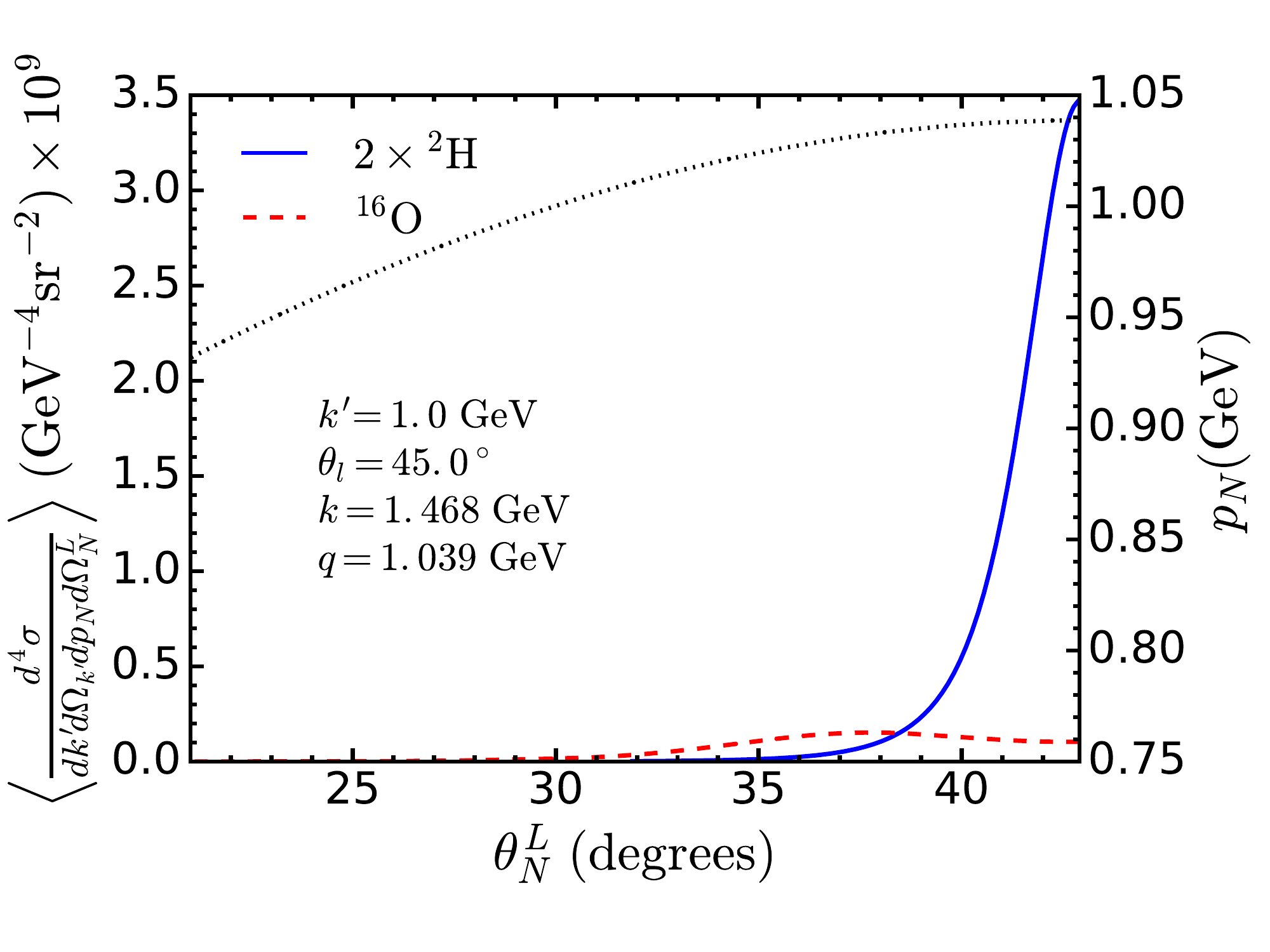}\includegraphics[height=2in]{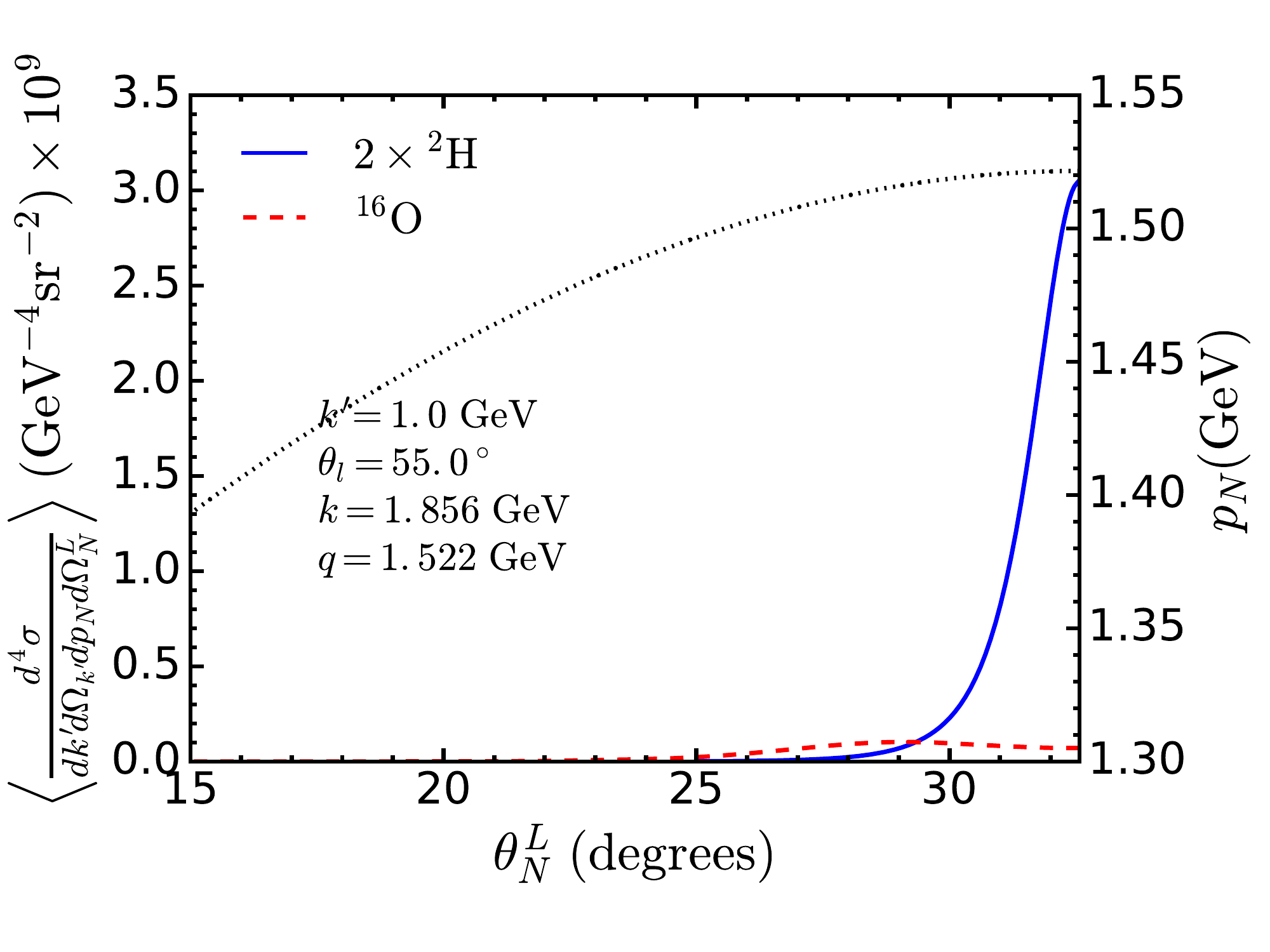}}
	\centerline{\includegraphics[height=2in]{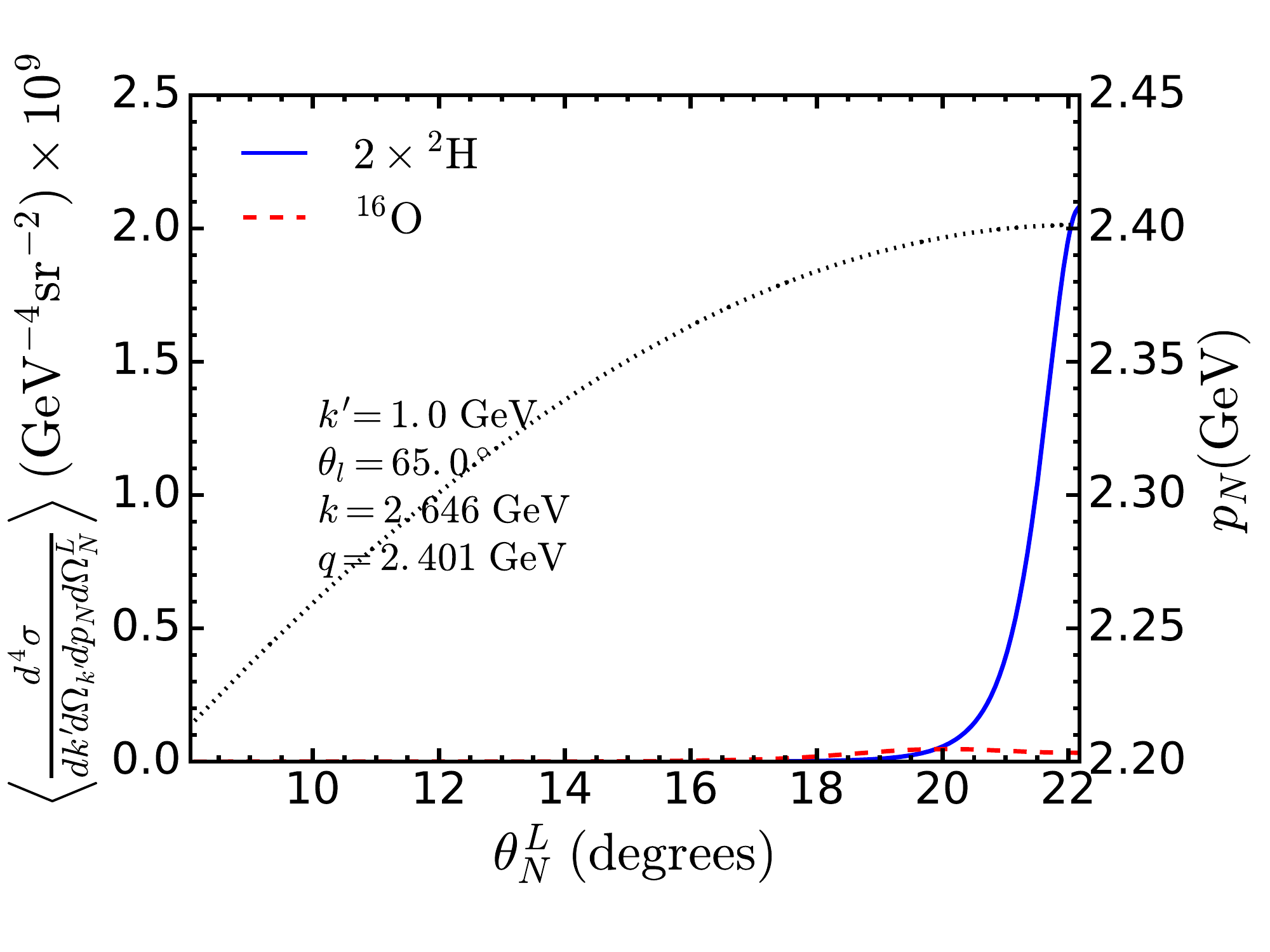}\includegraphics[height=2in]{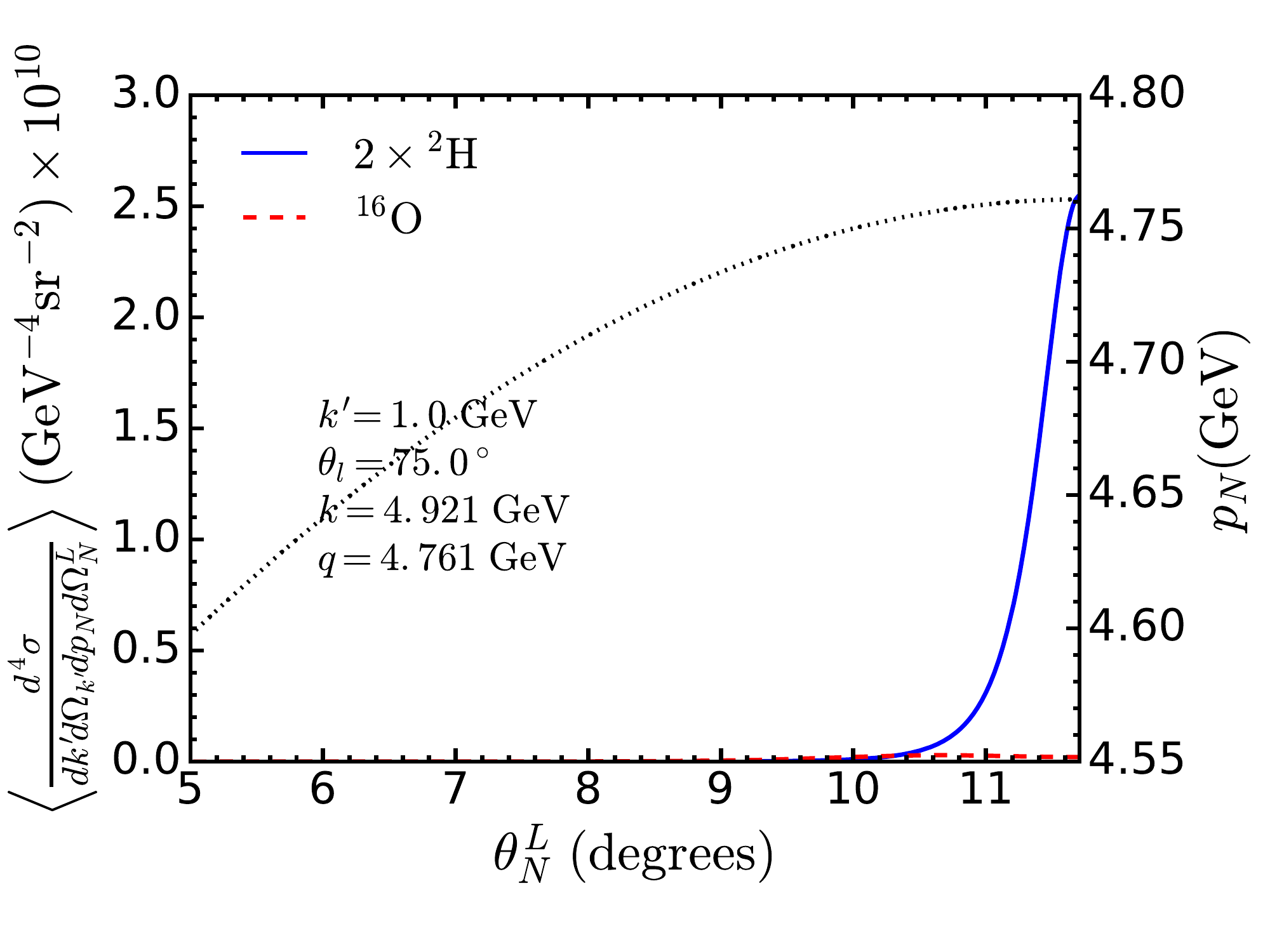}}
	\caption{(color online) Probability weighted cross sections for $k'=1$ GeV for various scattering angles $\theta_l$. The solid lines represent twice the deuteron cross section and the dashed lines are for the oxygen cross section versus $\theta^L_N$. The value of $p_N$ is represented by the dotted lines.}\label{fig:k_prime_1}
\end{figure}

\begin{figure}
	\centerline{\includegraphics[height=2in]{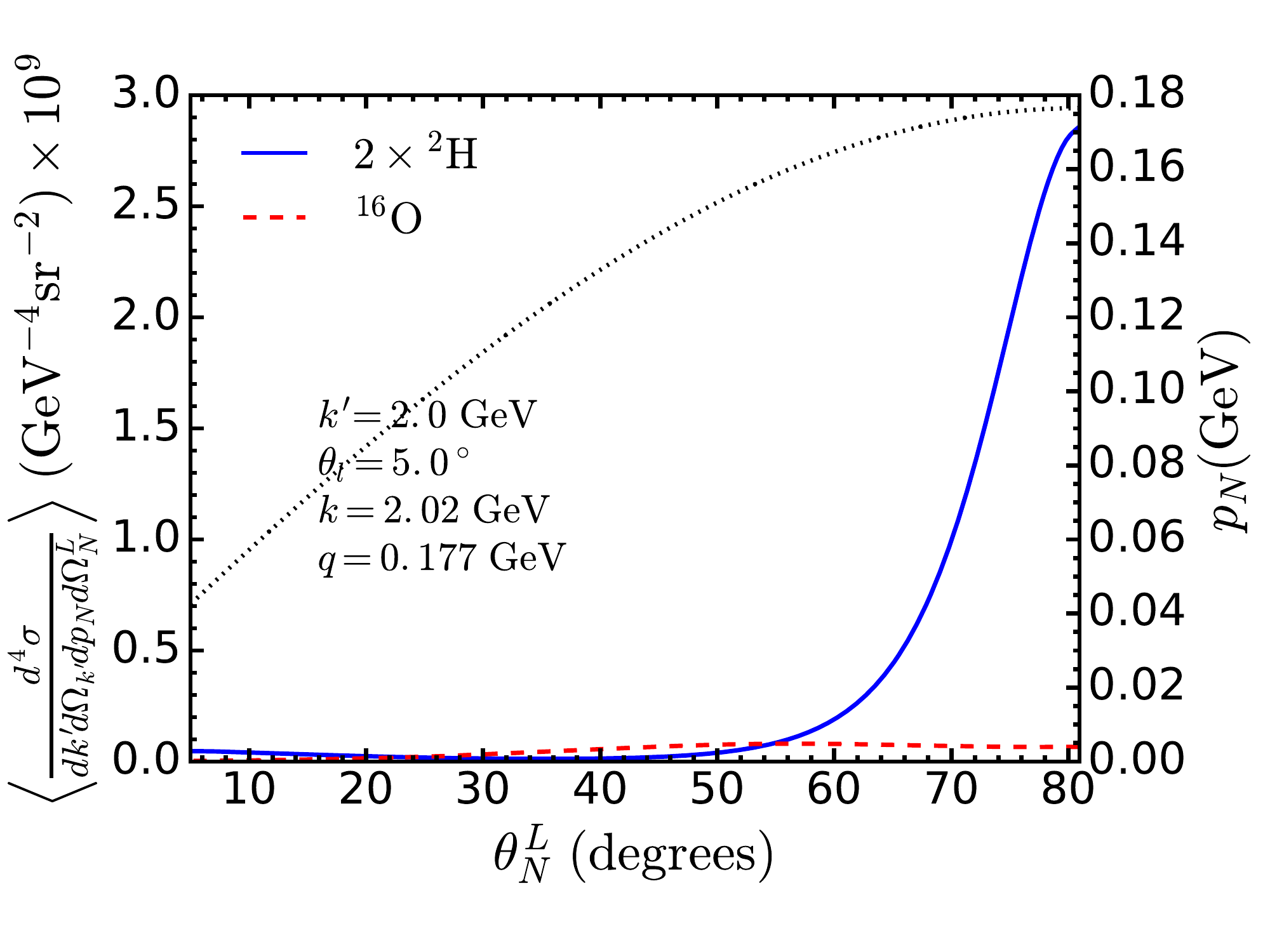}\includegraphics[height=2in]{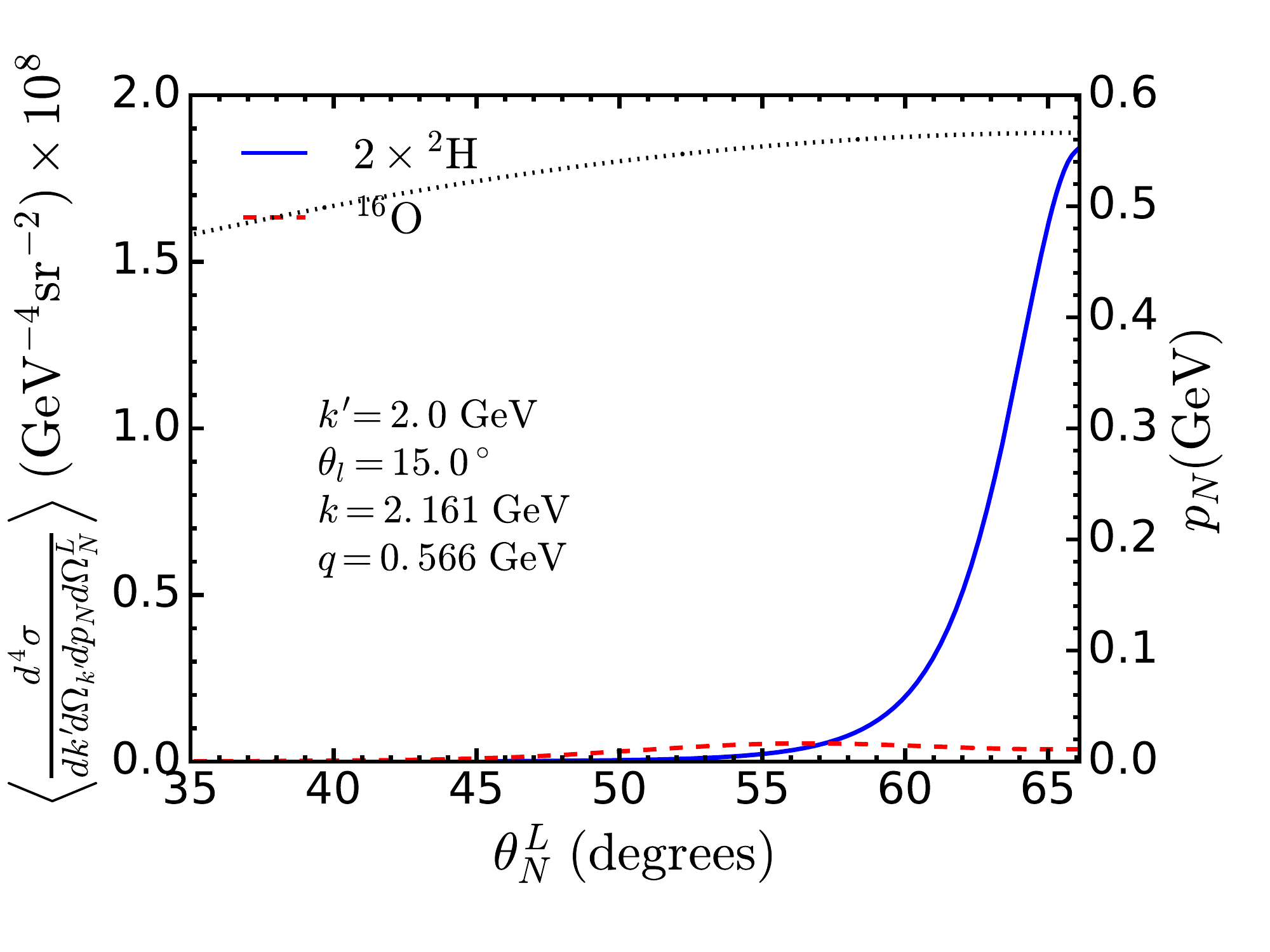}}
	\centerline{\includegraphics[height=2in]{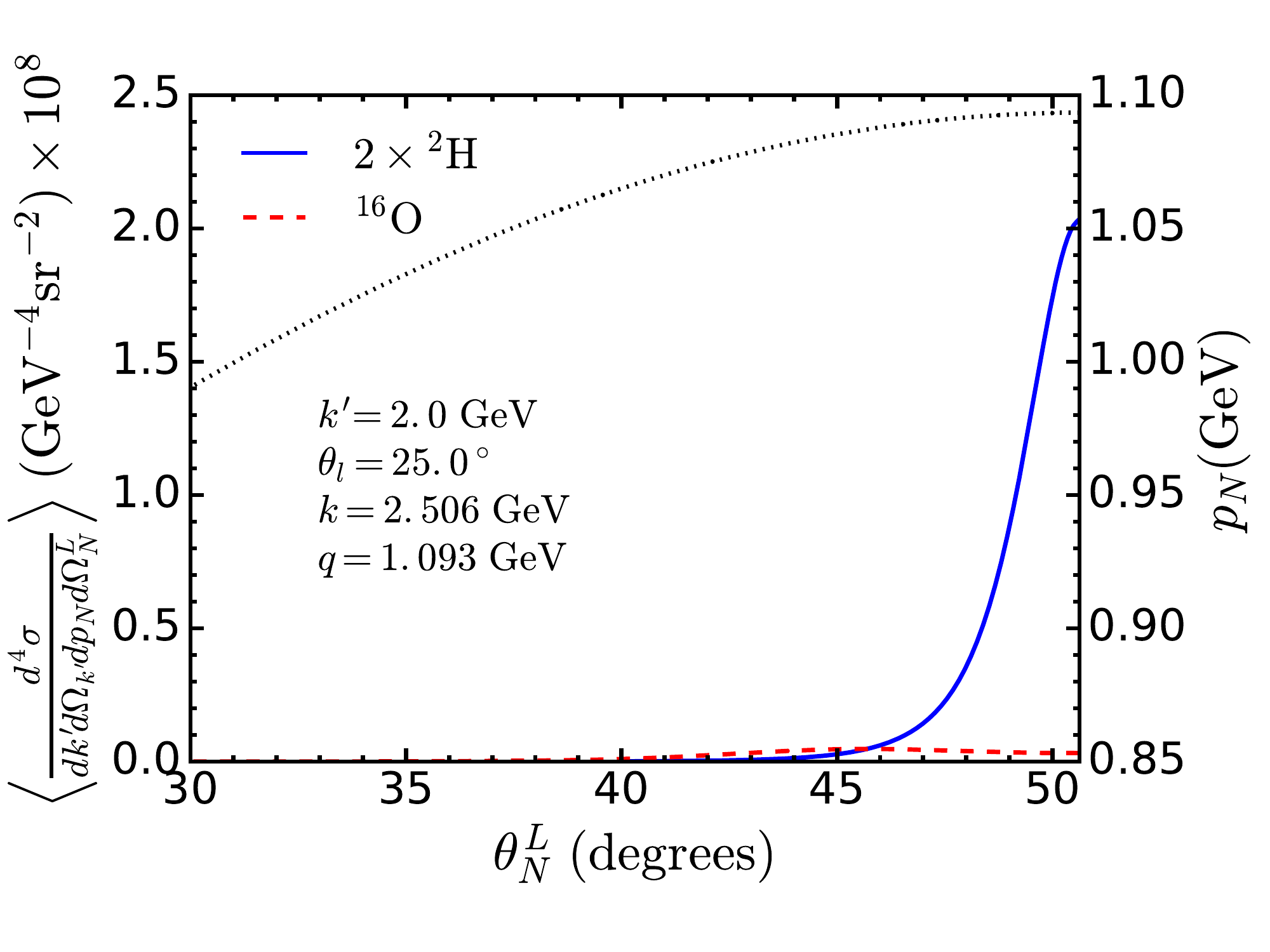}\includegraphics[height=2in]{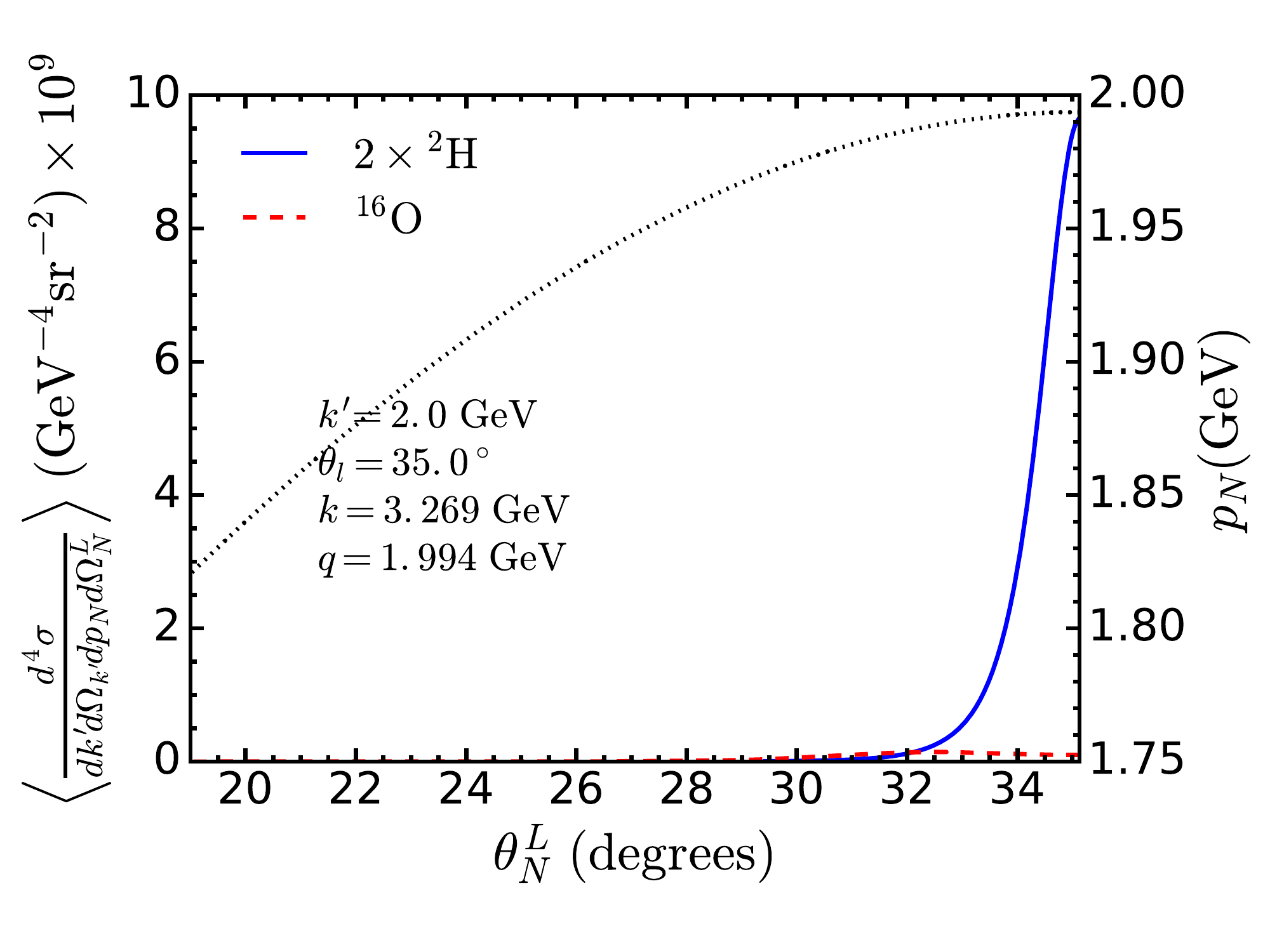}}
	\centerline{\includegraphics[height=2in]{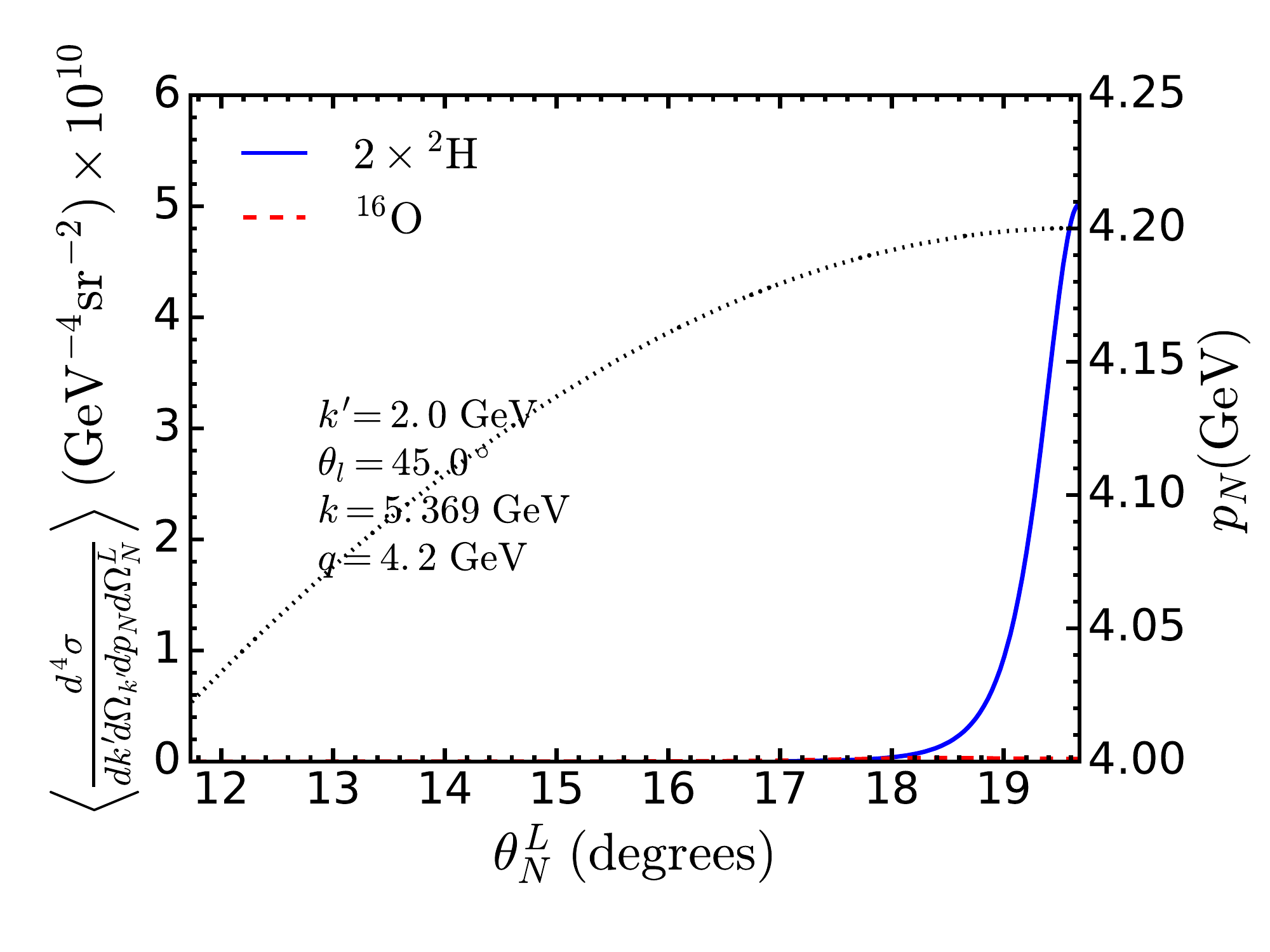}}
	\caption{(color online) As for Fig. \ref{fig:k_prime_1} but now for $k'=2$ GeV.  }\label{fig:k_prime_2}
\end{figure}

\begin{figure}
	\centerline{\includegraphics[height=2in]{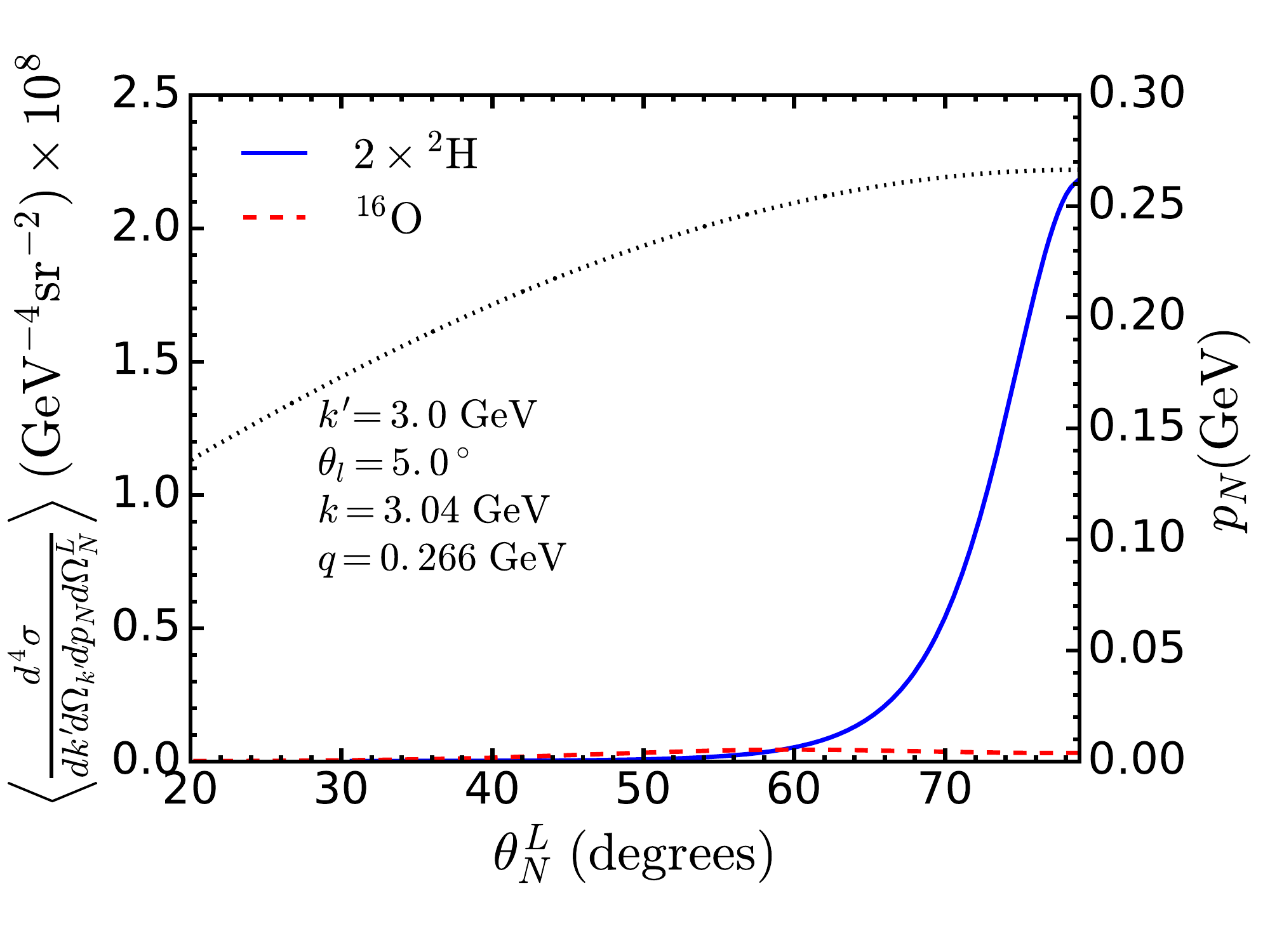}\includegraphics[height=2in]{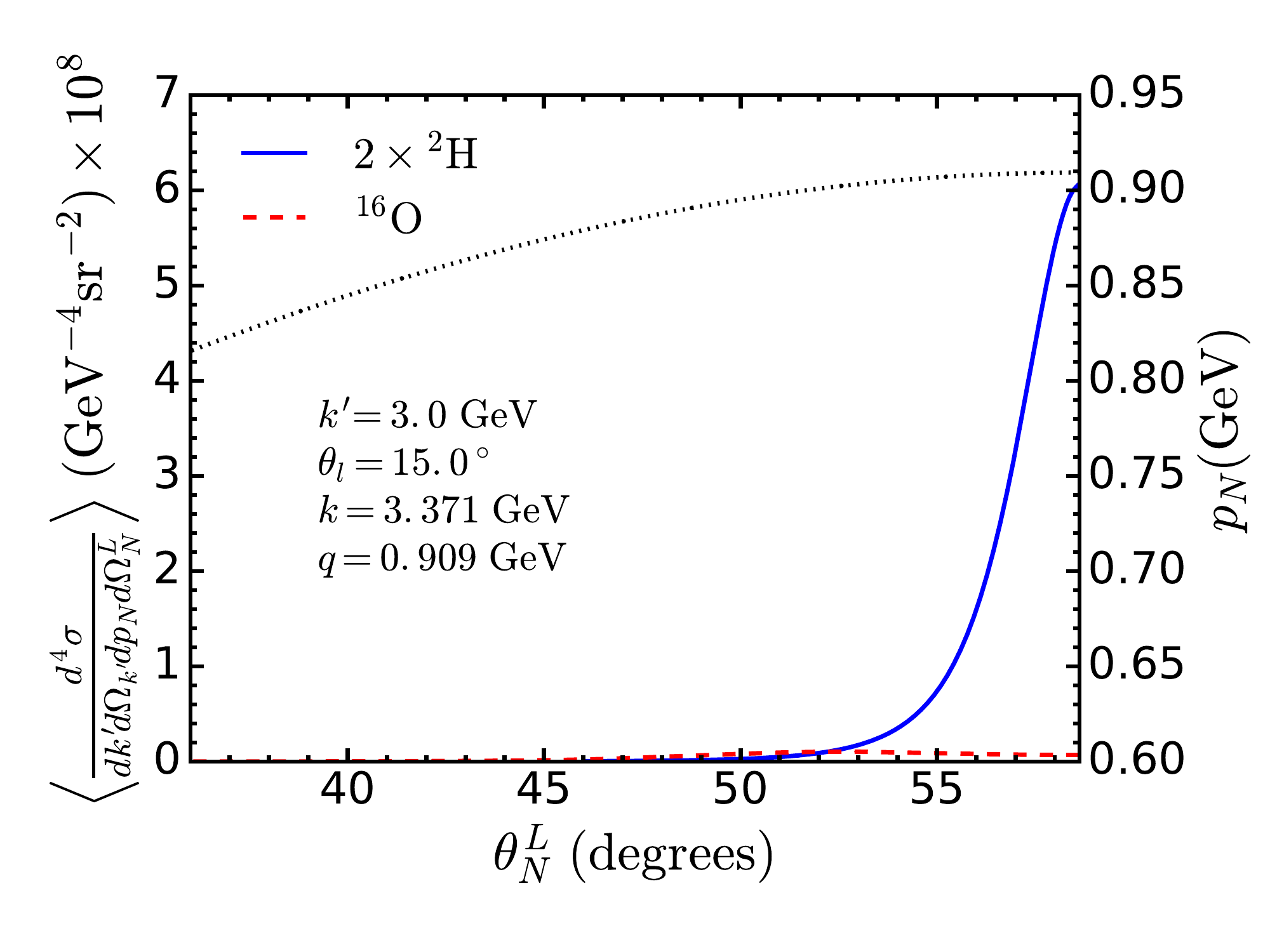}}
	\centerline{\includegraphics[height=2in]{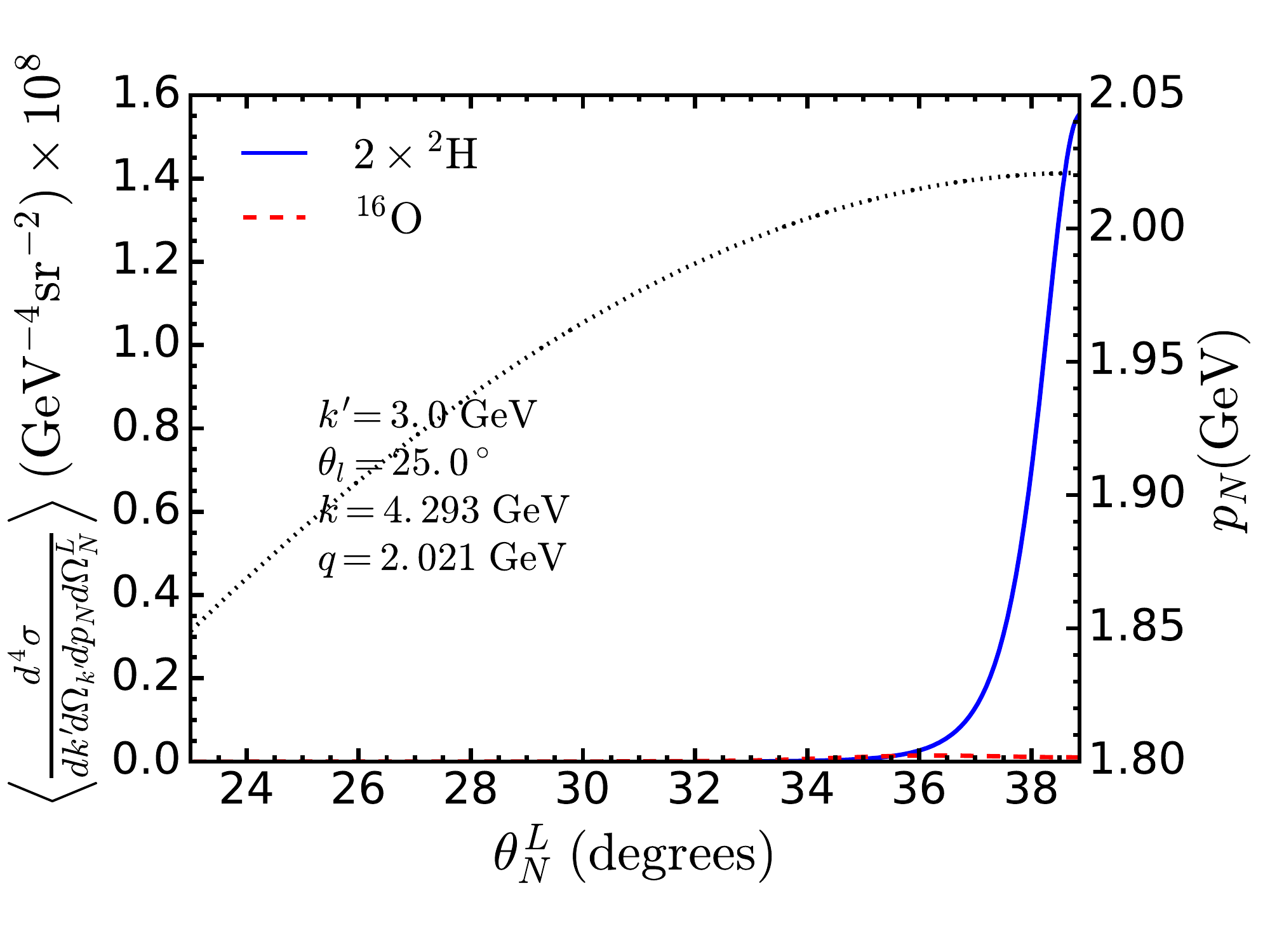}}
	\caption{(color online)  As for Fig. \ref{fig:k_prime_1} but now for $k'=3$ GeV.   }\label{fig:k_prime_3}
\end{figure}

Figures \ref{fig:k_prime_1}, \ref{fig:k_prime_2} and \ref{fig:k_prime_3} show cross sections for $^2\mathrm{H}$ and $^{16}\mathrm{O}$ for $k'=1,2\ \mathrm{and}\ 3\,\mathrm{GeV}$ respectively, as a function of the polar angle of the detected proton $\theta_N^L$, for a variety of lepton scattering angles subject to the $y=0$ constraint (see Sect.~\ref{sec:deut}) and the restriction required by Eq.~(\ref{eq:theta_l_max}). For each scattering angle, the values of the incident neutrino energy $k$ and the momentum transfer $q$ are given for the deuteron. For oxygen these quantities cover a range of values due to their dependence on the invariant mass $W_{A-1}$ which is integrated over to the  semi-inclusive cross section. For completeness, each figure contains the momentum of the detected proton $p_N$ as a function of $\theta_N^L$ with values given by the right-hand scale. Since for $^2\mathrm{H}_2{}^{16}\mathrm{O}$ there are two deuterium nuclei for each oxygen nucleus, the cross sections for deuterium are multiplied by a factor of 2. 
In all cases the maximum value of the oxygen cross section is at most one tenth of the deuterium cross section at its maximum value with the relative size decreasing for increased muon energy and scattering angle. It should be remembered, however, that these cross sections are evaluated and kinematics chosen to maximize the contribution of deuterium. 

\begin{figure}
	\centerline{\includegraphics[height=3in]{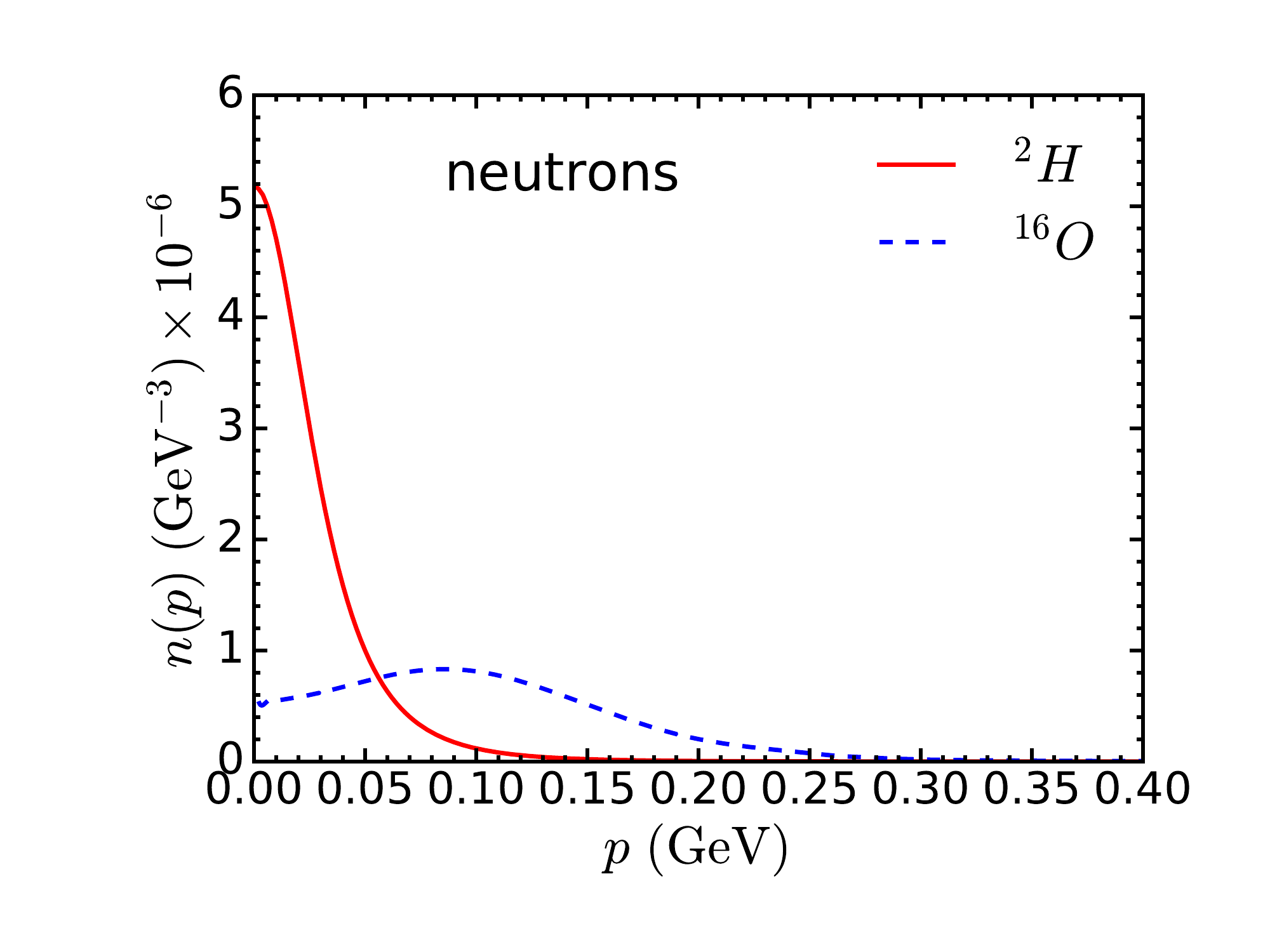}}
	\caption{(color online) Neutron momentum distributions for $^2$H (solid line) and $^{16}$O (dashed line).}\label{fig:momentum_dist}
\end{figure}

\begin{figure}
	\centerline{\includegraphics[height=3in]{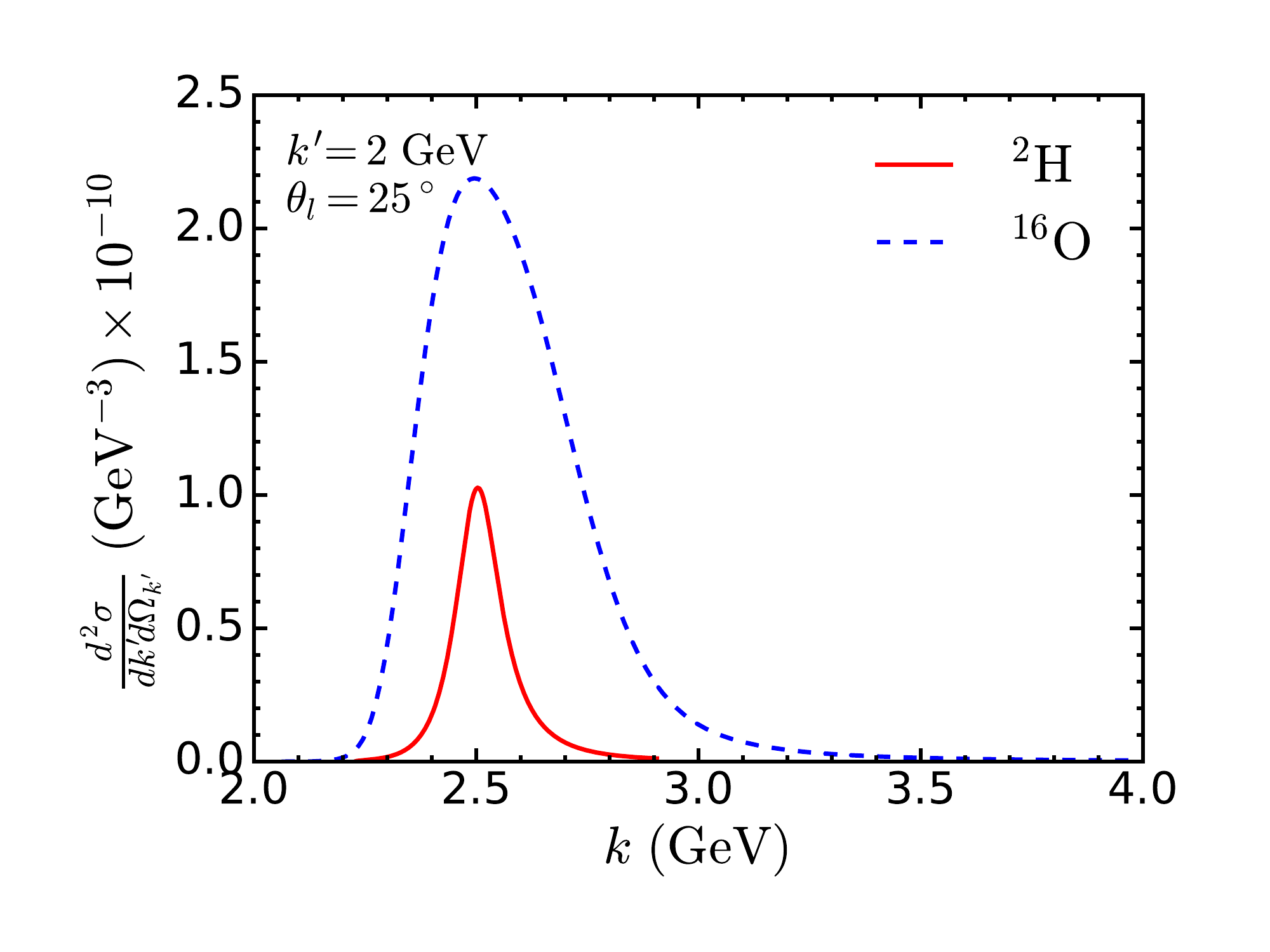}}
	\caption{(color online) Inclusive CC$\nu$ cross sections for $^2$H (solid line) and $^{16}$O (dashed line).}\label{fig:inclusive}
\end{figure}

The size of the deuterium cross sections relative to those of oxygen may still seem rather startling. The explanation for this is straightforward.  The semi-inclusive cross sections are roughly proportional to the neutron momentum distributions for the two nuclei as shown in Fig \ref{fig:momentum_dist}. Note that the maximum value of the deuterium momentum distribution is roughly five times as large as that for oxygen.  Given that there are two deuterium nuclei for each oxygen nucleus, this difference in the peak values of the deuterium and oxygen momentum distributions explains the difference in the size of the cross section shown above. 
Figure \ref{fig:inclusive} shows the inclusive cross sections for deuterium and oxygen as a function of the incident neutrino momentum. This shows that integrating over all possible values of proton three-momentum results in a much larger and broader quasielastic peak for oxygen than for deuterium, as should be expected. This indicates that the unconstrained semi-inclusive cross section is distributed over a much larger region of phase space than that for deuterium.

\section{Conclusions}\label{sec:concl}

The study presented in this paper of the semi-inclusive charge-changing neutrino reaction $(\nu_{\mu},\mu^- p)$ on a target of heavy water (D$_2$O) indicates that by careful choice of muon and proton three-momenta it is theoretically possible to separate deuterium events from those for oxygen. While naive considerations such as simply counting the number of neutrons provided by the two nuclei, namely, two for the two deuterium nuclei versus eight for the oxygen might lead one to expect that the latter will constitute a large background when the goal is to focus on events from the former, this proves not to be the case for events selected to favor the deuterium. As discussed in the previous section where results are given, this expectation is not necessarily the case:  the spectral function for deuterium is sharply peaked at small values of the missing momentum, whereas that for oxygen peaks at larger missing momenta where contributions from the 1p-shell are dominant and at low missing momenta but at higher missing energies where the 1s-shell contributions occur. Furthermore, these contributions to the oxygen spectral function are spread much more widely in missing momentum than the corresponding sharply peaked ones for deuterium, roughly by the factor of four obtained by forming the ratio of the Fermi momenta for the two nuclei, namely 55 MeV/c for deuterium and 230 MeV/c for oxygen.
In passing we note that the high missing energy/missing momentum region, while contributing perhaps 20\% to the inclusive cross section, is essentially irrelevant for the semi-inclusive cross section as the strength there is very broadly distributed and little is picked up when only limited parts of the ``nuclear landscape'' of the spectral function for a nucleus like oxygen are involved. 

In summary, from this theoretical study it appears that targets such as heavy water or deuterated methane containing significant amounts of deuterium together with light nuclei such as oxygen or carbon have the potential to provide unique information for studies of charge-changing neutrino reactions. Upon isolating the deuterium events using semi-inclusive reactions the kinematics alone will yield the incident neutrino energy on an event-by-event basis. Moreover, the cross section for such reactions on deuterium are arguably the best known throughout the periodic table even at quite high energies where relativistic modeling of the type used in the present work is undertaken. This being the case, such measurements hold the promise of determining the incident neutrino flux, thereby providing a very high-quality calibration of other existing or planned near detectors for neutrino oscillation experiments. Additionally, the fact that the nuclear structure issues are so well under control for the case of deuterium means that measurements of this type could serve in determining other aspects of the reaction, for instance, yielding new insights into the nature of the isovector axial-vector form factor of the nucleon. The issue now is an experimental one: can a practical target/detector of heavy water be realized? How are the protons in the final state to be detected? Can layers of (normal, un-deuterated) scintillator be used, as some have suggested, or are there other techniques to employ? Also: what is the optimal oscillation experiment using heavy water? While a near detector of heavy water appears worth contemplating, a far detector would be more challenging. Perhaps this last issue should be viewed in reverse, starting with the largest practical heavy water detector, then using the cross section to find how far from the neutrino source it could be placed, and then, finally, determining from the ``sweet spot'' for oscillation studies what beam energy is appropriate.

\appendix

\section{Off-shell single-nucleon response functions}\label{ap:offshell}

\begin{align}
8 m_N^4\widetilde{w}^{VV(I)}_{CC}=&4 E_p^2 \left(4 F_1^2(|Q|^2)
m_N^2+F^2_2(|Q|^2)
|Q|^2\right)+4 E_p \omega 
\left(4 F_1^2(|Q|^2)
m_N^2+F^2_2(|Q|^2)
|Q|^2\right)\nonumber\\
&-4 F_1^2(|Q|^2)
m_N^2 |Q|^2-8 F_1(|Q|^2)
F_2(|Q|^2) m_N^2 \left(\omega
^2+|Q|^2\right)\nonumber\\
&+F^2_2(|Q|^2)
\left(\omega ^2 |Q|^2-4 m_N^2
\left(\omega ^2+|Q|^2\right)\right)\nonumber\\
&-2\delta (2 E_p+\omega ) \left(F^2_2(|Q|^2)
\left(2 E_p \omega +\omega
^2-|Q|^2\right)-4 F_1^2(|Q|^2)
m_N^2\right)\nonumber\\
&+\delta^2\left(-4  E_p^2 F^2_2(|Q|^2)-12 E_p
F^2_2(|Q|^2) \omega +4 F_1^2(|Q|^2)
m_N^2+F^2_2(|Q|^2)\right.\nonumber\\
&\left.\times\left(|Q|^2-5 \omega ^2\right)\right)-4\delta^3 F^2_2(|Q|^2) (E_p+\omega )-\delta^4F^2_2(|Q|^2)\\
8 m_N^4\widetilde{w}^{AA(I)}_{CC}=&16 E_p^2 G^2_A(|Q|^2)
m_N^2+16 E_p G^2_A(|Q|^2)
m_N^2 \omega -4 G^2_A(|Q|^2)
m_N^2 \left(4
m_N^2+|Q|^2\right)\nonumber\\
&-8
G_A(|Q|^2) G_P(|Q|^2) m_N^2 \omega
^2+G^2_P(|Q|^2) \omega ^2 |Q|^2 \nonumber\\
&+\delta\left(16 E_p G^2_A(|Q|^2) m_N^2+8
G^2_A(|Q|^2) m_N^2 \omega\right.\nonumber\\
&\left. -8
G_A(|Q|^2) G_P(|Q|^2) m_N^2 \omega
-2 G^2_P(|Q|^2) \omega ^3\right)\nonumber\\
& +\delta^2\left(4 G^2_A(|Q|^2) m_N^2-G^2_P(|Q|^2)
\omega ^2\right)\\
8 m_N^4\widetilde{w}^{VV(I)}_{CL}=&2 E_p (2 p_\parallel+q) \left(4
F_1^2(|Q|^2) m_N^2+F^2_2(|Q|^2)
|Q|^2\right)+\omega  \left(8
F_1^2(|Q|^2) m_N^2 p_\parallel\right.\nonumber\\
&\left. -8
F_1(|Q|^2) F_2(|Q|^2) m_N^2
q+F^2_2(|Q|^2) \left(-4 m_N^2
q+2 p_\parallel |Q|^2+q
|Q|^2\right)\right)\nonumber\\
&\delta\left[\left(F^2_2(|Q|^2) \left(-\left(4 E_p^2
q+E_p \omega  (4 p_\parallel+6
q)+2 \omega ^2 (p_\parallel+q)-|Q|^2
(2 p_\parallel+q)\right)\right)\right.\right.\nonumber\\
&\left.\left.+8
F_1^2(|Q|^2) m_N^2 p_\parallel-4
F_1(|Q|^2) F_2(|Q|^2) m_N^2 q\right)\right]\nonumber\\
&-\delta^2F^2_2(|Q|^2) (4 E_p q+2 \omega 
p_\parallel+3 \omega  q)\nonumber\\
&-\delta^3F^2_2(|Q|^2) q\\
8 m_N^4\widetilde{w}^{AA(I)}_{CL}=&8 E_p G^2_A(|Q|^2) m_N^2 (2
p_\parallel+q)+\omega  \left(8
G^2_A(|Q|^2) m_N^2 p_\parallel-8
G_A(|Q|^2) G_P(|Q|^2) m_N^2
q\right.\nonumber\\
&\left.+G^2_P(|Q|^2) q |Q|^2\right)\nonumber\\
&\delta\left(8 G^2_A(|Q|^2) m_N^2 p_\parallel-4
G_A(|Q|^2) G_P(|Q|^2) m_N^2 q-2
G^2_P(|Q|^2) \omega ^2 q\right)\nonumber\\
&-\delta^2G^2_P(|Q|^2) \omega  q\\
8 m_N^4\widetilde{w}^{VV(I)}_{LL}=&16 F_1^2(|Q|^2) m_N^2 p_\parallel
(p_\parallel+q)-8 F_1(|Q|^2) F_2(|Q|^2)
m_N^2 q^2\nonumber\\
&+F^2_2(|Q|^2)
\left(|Q|^2 (2 p_\parallel+q)^2-4
m_N^2 q^2\right)-2\delta F^2_2(|Q|^2) q (2 E_p+\omega ) (2
p_\parallel+q)\nonumber\\
&-\delta^2 F^2_2(|Q|^2) q (4 p_\parallel+q)\\
8 m_N^4\widetilde{w}^{AA(I)}_{LL}=&16 G^2_A(|Q|^2) m_N^2 p_\parallel
(p_\parallel+q)-8 G_A(|Q|^2) G_P(|Q|^2)
m_N^2 q^2+G^2_P(|Q|^2) q^2
|Q|^2\nonumber\\
&-2\delta G^2_P(|Q|^2) \omega  q^2-\delta^2G^2_P(|Q|^2) q^2\\
8 m_N^4\widetilde{w}^{VV(I)}_{T}=&4 (4 F_1(|Q|^2) F_2(|Q|^2) m_N^2 |Q^2| + F_2^2(|Q|^2) (2 m_N^2 + p_\perp^2) |Q^2| \nonumber\\
&+ 
2 F_1^2(|Q|^2) m_N^2 (2 p_\perp^2 + |Q^2|))-16\delta F_1(|Q|^2) m_N^2 \omega\nonumber\\
& +\delta^2 \left[8 E_p^2 F^2_2(|Q|^2)+8 E_p
F^2_2(|Q|^2) \omega -8 F_1^2(|Q|^2)
m_N^2\right.\nonumber\\
&\left.-2 F^2_2(|Q|^2) |Q|^2
(F_1(|Q|^2)+F_2(|Q|^2))\right]+4\delta^3 F^2_2(|Q|^2) (2 E_p+\omega ) \nonumber\\
&+2\delta^4 F^2_2(|Q|^2)\\
8 m_N^4\widetilde{w}^{AA(I)}_{T}=&8 G^2_A(|Q|^2) m_N^2 \left(4
m_N^2+2
p_\perp^2+|Q|^2\right)-16\delta G^2_A(|Q|^2) m_N^2 \omega\nonumber\\
&-8\delta^2 G^2_A(|Q|^2) m_N^2\\
8 m_N^4\widetilde{w}^{VV(I)}_{TT}=&-4 p_\perp^2 \left(4 F_1^2(|Q|^2)
m_N^2+F^2_2(|Q|^2)
|Q|^2\right)\\
8 m_N^4\widetilde{w}^{AA(I)}_{TT}=&-16 G^2_A(|Q|^2) m_N^2
p_\perp^2\\
8 m_N^4\widetilde{w}^{VV(I)}_{TC}=&4 \sqrt{2} p_\perp (2 E_p+\omega
) \left(4 F_1^2(|Q|^2)
m_N^2+F^2_2(|Q|^2)
|Q|^2\right)\nonumber\\
&+4 \sqrt{2}\delta p_\perp
\left(F^2_2(|Q|^2) \left(-2 E_p
\omega -\omega ^2+|Q|^2\right)+4
F_1^2(|Q|^2) m_N^2\right)\nonumber\\
&-4\delta^2 \sqrt{2} F^2_2(|Q|^2) \omega 
p_\perp\\
8 m_N^4\widetilde{w}^{AA(I)}_{TC}=&16 \sqrt{2} G^2_A(|Q|^2) m_N^2
p_\perp (2 E_p+\omega )+16 \sqrt{2}\delta G^2_A(|Q|^2) m_N^2
p_\perp\\
8 m_N^4\widetilde{w}^{VV(I)}_{TL}=&4 \sqrt{2} p_\perp (2
p_\parallel+q) \left(4 F_1^2(|Q|^2)
m_N^2+F^2_2(|Q|^2)
|Q|^2\right)\nonumber\\
&-4 \sqrt{2}\delta F^2_2(|Q|^2) p_\perp q
(2 E_p+\omega )-4 \sqrt{2}\delta^2 F^2_2(|Q|^2) p_\perp q\\
8 m_N^4\widetilde{w}^{AA(I)}_{TL}=&16 \sqrt{2} G^2_A(|Q|^2) m_N^2
p_\perp (2 p_\parallel+q)\\
8 m_N^4\widetilde{w}^{VA(I)}_{T'}=&-32 G_A(|Q|^2) m_N^2
(F_1(|Q|^2)+F_2(|Q|^2)) (\omega 
p_\parallel-E_p q)\nonumber\\
&-16\delta G_A(|Q|^2) m_N^2 (2 F_1(|Q|^2)
p_\parallel-F_2(|Q|^2) q)\\
8 m_N^4\widetilde{w}^{VA(I)}_{TC'}=&-32 \sqrt{2} G_A(|Q|^2) m_N^2
p_\perp q (F_1(|Q|^2)+F_2(|Q|^2))\nonumber\\
&-4 \sqrt{2}\delta F_2(|Q|^2) G_P(|Q|^2) \omega 
p_\perp q\\
8 m_N^4\widetilde{w}^{VA(I)}_{TL'}=&-32 \sqrt{2} G_A(|Q|^2) m_N^2 \omega 
p_\perp (F_1(|Q|^2)+F_2(|Q|^2))\nonumber\\
&-4 \sqrt{2}\delta p_\perp \left(8
F_1(|Q|^2) G_A(|Q|^2)
m_N^2-F_2(|Q|^2) G_P(|Q|^2)
q^2\right)\,,
\end{align}
where
\begin{align}
p_\parallel&=\frac{\bm{p}\cdot\bm{q}}{q}\\
p_\perp&=\frac{|\bm{p}\times\bm{q}|}{q}\\
E_p&=\sqrt{p^2+m_N^2}\\
|Q^2|&=q^2-\omega^2\,.
\end{align}
The isovector electromagnetic form factors $F_1$ and $F_2$ are from \cite{FF_GK_05_2_Lomon_02,FF_GK05_1_Lomon_06} and the weak form factors $G_A$ and $G_P$ are simple dipole forms as used in \cite{deut}.

\begin{acknowledgments}
This work has been supported in part by the Office of Nuclear Physics of the US Department of Energy under Grant Contract DE-FG02-94ER40818 (T. W. D.), by the US Department of Energy under Contract No. DE-AC05-06OR23177, and by the U.S. Department of Energy cooperative research agreement DE-AC05-84ER40150 (J. W. V. O.). O. M. acknowledges support from a Marie Curie International Outgoing Fellowship within the European Union 7th Framework Programme under Grant Agreement PIOF-GA-2011-298364 (ELECTROWEAK), and from the MINECO project grant FIS2014-51971-P. The authors also thank Omar Benhar for kindly providing a model for the spectral function of ${}^{16}$O employed in this study.
\end{acknowledgments}

\bibliography{neutrino}




\end{document}